\documentclass[11pt,a4paper]{article}
\usepackage[inactive, tightpage]{preview}
\usepackage{jheppub}
\usepackage{amsthm}
\usepackage[T1]{fontenc}
\usepackage[utf8]{inputenc}
\usepackage{mathtools}   
\usepackage{amsfonts}
\usepackage{physics}
\usepackage{mathrsfs}
\usepackage{natbib}
\usepackage{comment}
\usepackage{tikz, pgfplots} 
\usetikzlibrary{patterns, calc, patterns, angles, quotes, decorations, shadows}
\usepackage{enumitem}
\usepackage{subcaption}
\usepackage[font={small}]{caption}
\usepackage{graphicx}
\usepackage{amssymb}
\usepackage{microtype}

\usepackage{microtype}
\usepackage{hyperref}

\definecolor{darkmidnightblue}{rgb}{0.0, 0.2, 0.4}
\hypersetup{
    colorlinks = true, 
    linkcolor=darkmidnightblue,
    citecolor=darkmidnightblue
}

\usepackage{cleveref}

\usepackage{booktabs}
\usepackage{pdflscape}
\usepackage{afterpage}
\usepackage{flafter}
\usepackage{rotating}
\usepackage{fancyhdr}
\usepackage{slashed}
\fancypagestyle{lscape}{%
\fancyhf{} 
\fancyfoot[LE]{}
\fancyfoot[LO] {}
 
}

\newcommand{\ep}{\epsilon}

\newcommand{\C}{\mathbb{C}} 

\newcommand{\Hom}{\text{\Hom}}

\DeclareMathSymbol{:}{\mathord}{operators}{"3A}

\title{Boundaries \& Localisation with a Topological Twist}

\author[a,b]{Samuel Crew,}
\author[c,d]{Daniel Zhang,}
\author[e]{Boan Zhao}

\affiliation[a]{Faculty of Computer Science, Ruhr University Bochum, Germany.}

\affiliation[b]{Max Planck Institute for Security and Privacy, Bochum, Germany.}

\affiliation[c]{St John's College, University of Oxford, Oxford, United Kingdom}

\affiliation[d]{Mathematical Institute, University of Oxford, Oxford, United Kingdom}

\affiliation[e]{DAMTP, University of Cambridge, Cambridge, United Kingdom}

\emailAdd{samuel.crew@rub.de, daniel.zhang@sjc.ox.ac.uk, bz258@cam.ac.uk}

\abstract{
We study the partition functions of topologically twisted 3d $\mathcal{N}=2$ gauge theories on a hemisphere spacetime with boundary $HS^2 \times S^1$. We show that the partition function may be localised to either the Higgs branch or the Coulomb branch where the contributions to the path integral are vortex or monopole configurations respectively. Turning to $\mathcal{N}=4$ supersymmetry, we consider partition functions for exceptional Dirichlet boundary conditions that yield a complete set of `IR holomorphic blocks'. We demonstrate that these correspond to vertex functions: equivariant Euler characteristics of quasimap moduli spaces. In this context, we explore the geometric interpretation of both the Higgs and Coulomb branch localisation schemes in terms of the enumerative geometry of quasimaps and discuss the action of mirror symmetry. 
}

\theoremstyle{definition}

\begin{document}
\maketitle

\section{Introduction}

Supersymmetric indices are powerful tools to study supersymmetric quantum field theories. Localisation techniques allow us to explicitly compute these indices and provide connections between supersymmetric quantum field theories and moduli space geometry. In three dimensions, the Coulomb branch localisation of $\mathcal{N}=2$ topologically twisted theories on $S^2\times S^1$ was first considered in \cite{benini2015topologically} and extended to other closed Riemann surfaces in \cite{Benini:2016hjo, Closset_2016}. The study of Higgs branch localisation for 2d $\mathcal{N}=(2,2)$ theories on $S^2$ was initiated by \cite{benini2015partition, Doroud:2012xw} and later extended to $S^3$ partition functions and superconformal indices on $S^2 \times S^1$ in the works \cite{benini2014higgs,fujitsuka2014higgs}.

In this work we study the partial topological twist of 3d $\mathcal{N}=2$ theories on a spacetime with boundary $HS^2 \times S^1$, where $HS^2$ is the 2d hemisphere. Our first result is to formulate supersymmetric boundary conditions on $\partial(HS^2 \times S^1)=T^2$ for this spacetime and compute the Witten index via both Higgs and Coulomb branch localisation. In particular, we provide a formula for the index with Dirichlet-type boundary conditions for the $\mathcal{N}=2$ vector multiplet. This provides a UV derivation of the holomorphic blocks \cite{beem2014holomorphic}, which are defined to be the partition function of partially topologically twisted theories on $S^1$ times an infinite cigar and typically computed with effective quantum mechanics in the IR.

With these results in hand, we proceed to the main object of interest in this work: the index for 3d $\mathcal{N}=4$ gauge theory with a topological twist preserving the Blau and Thomson \cite{blau1997aspects} supercharge $Q_B$. In the IR, this twist coincides with the Rozansky-Witten twist \cite{rozansky1997hyper} and is known as either the $B$ or $C$ twist in various parts of the literature. We study the index of a particular set of boundary conditions known as \textit{exceptional Dirichlet}. These $\mathcal{N}=(2,2)$ preserving boundary conditions, labelled $\{D_{\alpha}\}$, were first studied in \cite{bullimore2016boundaries} and were motivated by analogous boundary conditions for 2d $\mathcal{N}=(2,2)$ theories \cite{Hori:2000ck, Gaiotto:2015aoa}. Exceptional Dirichlet boundary conditions are defined with respect to a choice of chamber in the mass parameter space for the Higgs branch flavour symmetry and are in one-to-one correspondence with massive Higgs branch vacua of the theory. In the IR they are supported on holomorphic attracting Lagrangian submanifolds associated to such vacua and, in this way, are designed to mimic the presence of an isolated vacuum $\alpha$ at infinity for the purposes of BPS computations. They play an important role in 3d mirror symmetry \cite{bullimore2016boundaries} and the connections between equivariant elliptic cohomology and $\mathcal{N}=4$ gauge theories \cite{bullimore20223d,dedushenko2021interfaces}. 

The (twisted, flavoured) Witten index of this boundary condition may be computed with a path integral and counts states in the cohomology of one of the supercharges preserved by the topological twist
\begin{equation}
    \mathcal{Z}_{D_{\alpha}} = \Tr_{\mathcal{H}_{HS^2}}(-1)^F e^{-\beta H}q^J t^{\frac{R_H-R_C}{2}} x^{F_H} \xi^{F_C}.
\end{equation}
The index is graded by fermion number and global symmetries, which may be implemented by turning on background real masses and holonomies. The fugacities $t$, $x$ and $\xi$ here are exponentials of a supersymmetric complex combination $A_3^{(F)} + i m^{(F)}$ of holonomy and mass parameters for a background vector multiplet that weakly gauges the relevant flavour symmetry. Here, $R_H$ and $R_C$ are the generators of the Cartan of $SU(2)_H\times SU(2)_C$ $R$-symmetry, $F_H$ the Higgs branch flavour symmetries, and $F_C$ the topological symmetry.

In the work \cite{bullimore2021boundaries}, two of the authors studied the closely related half superconformal index. This may again be computed on the spacetime $HS^2\times S^1$ but now with the same rigid supersymmetry background as for the superconformal index on $S^2\times S^1$ and, in particular, is not topologically twisted. This index was computed for Neumann boundary conditions for the vector multiplet in \cite{yoshida2014localization}, and Dirichlet in \cite{bullimore2021boundaries}. In \textit{loc. cit.}, it was demonstrated that 
\begin{equation}
    \mathcal{Z}_{\text{SC}} = e^{\varphi} \mathcal{I},
\end{equation}
where $\mathcal{I}$ is a count of local operators supported on the boundary condition, studied in \cite{gadde2014walls, gadde2016fivebranes, dimofte2018dual} and $\varphi$ may be interpreted as a supersymmetric Casimir energy \cite{Bobev:2015kza} encoding boundary 't Hooft anomalies. 
As anticipated in \cite{bullimore2021boundaries}, $\mathcal{Z}_{\text{SC}}$ with $\{D_{\alpha}\}$ boundary conditions and $\mathcal{Z}_{D_{\alpha}}$ are identical up to a shift in fugacities $t \rightarrow tq^{-\frac{1}{2}}$. In this work, we provide a physical argument for this correspondence and verify it explicitly with a localisation computation.
The half superconformal indices were found to realise exactly the aforementioned holomorphic factorisation \cite{beem2014holomorphic} of various three-manifold partition functions of 3d $\mathcal{N}=4$ theories and the index we consider in this work naturally factorises the topologically $C$-twisted index \cite{crew2020factorisation, Cabo-Bizet:2016ars}.

Supersymmetric partition functions may be localised using different schemes giving rise to geometric dualities. The work of Bullimore \textit{et. al.} \cite{bullimore2019twisted} explores the geometric interpretation of the Higgs branch localisation scheme for theories on $\Sigma \times S^1$ with $\Sigma$ a closed Riemann surface. By introducing an additional $Q$-exact action to the Coulomb branch localising action, the authors localise to vortex configurations on $\Sigma$ which are given by solutions to:
\begin{equation}\label{eq:bpslocusintro}
\begin{aligned}
&\qquad\qquad  2iF_{z\bar{z}} = \tau-2\mu_{\mathbb{R}},  \quad \mu_{\mathbb{C}} = 0, \quad F_{z3} = 0, \\
&\qquad\qquad(\nabla_{\bar{z}} + iA_{\bar{z}})X = 0, \quad (D_3 + iA_3)X_i=0, \\
&\qquad\qquad(\nabla_{\bar{z}} - iA_{\bar{z}})Y = 0, \quad (D_3 -iA_3)Y_i =0,\\
&\qquad\quad \sigma \cdot X_i = 0, \quad \sigma \cdot Y_i = 0, \quad \phi_{\bar{z}}X_i = 0, \quad \phi_{\bar{z}}Y_i = 0,\\
&\qquad\qquad\qquad\qquad\quad \nabla_\mu\sigma = \nabla_\mu\phi_{\bar{z}} = 0.
\end{aligned}
\end{equation}
where $(X,Y)$ are hypermultiplet scalars and $(\sigma, \phi)$ are vector multiplet scalars. In the $C$-twist, the moduli space of solutions to the above equations has an algebraic description as the moduli space of quasimaps $\Sigma \rightarrow \mathcal{M}_H$ to the Higgs branch \cite{okounkov2015lectures} (in the $H$ or $A$ twist, one recovers the twisted quasimaps of \cite{kim2012stable}). The twisted index on $\Sigma \times S^1$ computes the equivariant Euler characteristic of this moduli space.

In this work we are concerned with both the Coulomb and Higgs branch localisation of topologically twisted theories instead on a spacetime with boundary, $HS^2\times S^1$. The Coulomb branch localisation proceeds similarly to \cite{benini2015topologically}, the BPS locus consists of monopole configurations compatible with the Dirichlet boundary condition for the gauge field. The Higgs branch localisation leads to the same BPS locus of vortices as \eqref{eq:bpslocusintro}. In both cases, the configurations are now subject to exceptional Dirichlet boundary conditions on the torus boundary. The result of both localisation procedures may be obtained from our formulae for the index for $\mathcal{N}=2$ theories, which are derived for multiplets in a generic flux $k$ background.

The result of the Coulomb branch localisation first depends on an additional parameter $s=\exp(i\beta(A_3+i\sigma)|_{\partial})$ which is the (exponentiated) constant value of the gauge field $A_3$ and real scalar $\sigma$ at the boundary. The exceptional Dirichlet boundary condition switches on boundary vevs corresponding to a vacuum configuration $\alpha$ for the chiral multiplets, and $s$ must consequently be specialised to a value dependent on $(x_i,t)$ in order to preserve supersymmetry. Equivalently, the total effective connection and mass of the chirals which attain boundary vevs must vanish. This substitution is the analog of the Chern root evaluation of equivariant $K$-theory classes for GIT quotients at torus fixed points. We find that:
\begin{equation}\label{eq:introindex}
    \mathcal{Z}^{\text{C.B.}}_{D_{\alpha}}(q,t,s=s_{\alpha}, x,\xi) = \mathcal{Z}_{D_{\alpha}}(q,t,x,\xi) = e^{\phi_{\alpha}}\mathcal{Z}^{\text{1-loop}}_{\alpha}\mathcal{Z}_{\alpha}^{\text{vortex}}(x,\xi,q,t),
\end{equation}
where $\mathcal{Z}_{D_\alpha}$ is the twisted index on the hemisphere, which we also obtain directly from Higgs branch localisation. In the above, $e^{\phi_{\alpha}}$ arises from classical Fayet-Iliopoulos action evaluated on the BPS locus, and effective Chern-Simons contributions arising from integrating out massive fermions. $\mathcal{Z}^{\text{1-loop}}_{\alpha}$ is a perturbative one-loop determinant, and $\mathcal{Z}_{\alpha}^{\text{vortex}}$ encodes contributions from vortices.

We then turn to one of the main conceptual results of this work, which is to explicitly relate the above partition function to the enumerative geometry of \textit{based} quasimaps $\textsf{QM}_{\alpha}$ from $\mathbb{P}^1 \rightarrow \mathcal{M}_H$ \cite{okounkov2015lectures}. This moduli space splits into components labelled by the degree $k$ of the $G$ bundle over $\mathbb{P}^1$, which we denote $\textsf{QM}^{k}_{\alpha}$. The central mathematical object is the \textit{vertex function}, which is the generating function of the (equivariant) Euler characteristics of the virtual structure sheaf over the different degree quasimap moduli spaces. 
\begin{equation}
    \mathsf{V}_{\alpha} = \sum_{k} \xi^{k} \chi_{\mathsf{T}}(\textsf{QM}^{k}_{\alpha}, \hat{\mathcal{O}}_{\text{vir.}}).
\end{equation}

We demonstrate that the index $\mathcal{Z}_{D_\alpha}$ computes $\mathsf{V}_{\alpha}$ up to a prefactor, to which we also give a geometric interpretation. The exceptional Dirichlet boundary condition is designed to mimic the presence of an isolated, massive vacuum at infinity, as such it is reasonable to expect that $\mathcal{Z}_{D_\alpha}$ is equal to the partition function on $\mathbb{R}^2 \times S^1$, with a vacuum at $\infty$, as it is computed on a topologically twisted background. Up to perturbative prefactors (which are familiar from say, the relationship between the Nekrasov partition function on $S^4$ \cite{Nekrasov:2002qd} and the Seiberg-Witten prepotential for theories on $\mathbb{R}^4$ \cite{Seiberg:1994rs}, see \textit{e.g.} \cite{Shadchin:2005mx}), one would expect this to correspond to a $\mathbb{P}^1$ partition function with the vacuum $\alpha$ imposed at a pole. The latter also localises to the same BPS equations as \eqref{eq:bpslocusintro}, but now on $\mathbb{P}^1$ but with a condition at the south pole. It is expected therefore that this $\mathbb{P}^1$ partition function, which we call $\mathcal{Z}_{\alpha}$, yields the vertex function. We perform an explicit Higgs branch localisation on the twisted $\mathbb{P}^1$ index, with a vacuum condition at the south pole, to elucidate this correspondence. 

The Higgs branch localisation scheme recovers the (virtual) equivariant localisation formula for the vertex function. On the other hand the index $\mathcal{Z}_{D_{\alpha}}$ on $HS^2\times S^1$ may be localised to the Coulomb branch. We argue that in this case one recovers a $q$-Jackson integral formula for the index of the form
\begin{equation}
    \mathcal{Z}_{D_\alpha}(x_i,\xi, t,q) = \int_{0}^{s(\alpha)} d_q s \, e^{-\phi_0(s,\xi)} \hat{a} \left[ \frac{1-t^{-1}q}{1-q} \left( Q^{+} + \bar{Q}^{0} - \mu_{\mathbb{C}} \right) \right] \,,
\end{equation}
where $Q^+$, $\bar{Q}^0$ and $\mu_{\mathbb{C}}$ denote certain weight subspaces of $T \mathcal{M}_H$ determined by the exceptional Dirichlet boundary condition $D_{\alpha}$.

We show that this localisation formula naturally gives rise to a chamber dependent perturbative contribution. This normalisation is often added to vertex functions by hand in order to show that they satisfy appropriate $q$-difference equations in quantum $K$ theory \cite{dinkins2022exotic}. The Coulomb branch localisation hence provides a physical meaning to these perturbative terms. In a similar vein, we briefly clarify the physical interpretation of the elliptic stable envelopes \cite{Aganagic:2016jmx} in relating the vertex functions for symplectic dual varieties. This is intimately related to the action of mirror symmetry on the exceptional Dirichlet boundary conditions of a pair of symplectic (mirror) dual varieties, which yields the enriched Neumann boundary conditions introduced in \cite{bullimore2016boundaries,bullimore20223d}.

\paragraph{Outline. }The paper is organised as follows. In section \ref{sec:background} we fix our notation for 3d topologically twisted supersymmetric gauge theories and then review the relevant aspects of 3d $\mathcal{N}=4$ supersymmetric theories including their Higgs branch geometry. In section \ref{sec:twistedtheories} we introduce boundary conditions for twisted theories, and define the twisted index on $HS^2\times S^1$. We also discuss the relation with the half superconformal index. In section \ref{sec:localisation} we describe the Higgs and Coulomb branch localisation schemes, and derive the corresponding one-loop determinants, which hold also for $\mathcal{N}=2$ theories. We compute the index for  $\mathcal{N}=4$ QED with $N$ fundamental hypermultiplets as an illustrative example. In section \ref{sec:quasimap} we elucidate the equivalence between the twisted index and the vertex function, establishing correspondences between both perturbative and vortex contributions to the index with objects in the geometry of based quasimap moduli spaces. Finally, we comment on the physical interpretation of the role of elliptic stable envelopes \cite{Aganagic:2016jmx} in the pole subtraction formula, as the the action of mirror symmetry on the indices we consider. We include various technical appendices required for the localisation computations.

\paragraph{Further directions.}

This work primarily considers abelian theories, for the sake of simplicity. It would be interesting to explore non-abelian examples. More interestingly, from a physical perspective the hemisphere geometry $S^1\times HS^2$ appears rather naturally: it allows the prescription of a boundary condition, it is intuitively the correct perspective for holomorphic factorisation, and it is related to a count of boundary local operators via the state-operator correspondence. For example, as already demonstrated in \cite{bullimore2021boundaries}, localising gives the correct perturbative contributions to factorise three-manifold partition functions. On the other hand, geometrically it is less clear how to develop directly a deformation-obstruction theory for vortices on a hemisphere with a finite-distance boundary condition. It would be interesting to explore further the virtual localisation structure implied by the vortex equations. Finally. it would be interesting to investigate the $H$-twist in the presence of boundaries; one should recover a based version of the twisted quasimaps of \cite{kim2012stable}.

\paragraph{Note.} Whilst preparing this manuscript we became aware of overlapping work by Mykola Dedushenko and Nikita Nekrasov. We thank them for interesting discussions and for agreeing to coordinate the submission.

\section{Background}\label{sec:background}

In this section we introduce our conventions for quasi-topologically twisted 3d $\mathcal{N}=2$ theories. We first write down the transformations and Lagrangian on a generic product space $\Sigma \times S^1$, where $\Sigma$ may have a non-trivial boundary. We denote the length of the time circle $S^1$ to be $\beta$. We will also, for simplicity, restrict to an abelian gauge group $G = U(1)^r$ throughout this work, although we expect the results to hold more generally.

\subsection{\texorpdfstring{$\mathcal{N}=2$}{} supersymmetry}\label{eq:susytransforms}

We now introduce $\mathcal{N}=2$ field content and Lagrangian for a topological twist along $\Sigma$ with respect to a $U(1)_R$ $R$-symmetry. An over-line on fermions indicates positive $U(1)_R$ charge. Fermions with positive $U(1)_R$-charge are twisted to $(0,1)$ forms or scalars.

We use $\nabla$ to denote the Levi-Civita connection on $\Sigma$ and $D = \nabla + iA$ to denote the gauge covariant derivative. We use conventions where $\nabla_z = \frac{1}{2}(\nabla_1 - i\nabla_2), \nabla_{\bar{z}} = \frac{1}{2}(\nabla_1 + i\nabla_2)$ and $1,2$ are the indices of an oriented orthonormal frame at a point on $\Sigma$ ($z$ is not a complex coordinate on $\Sigma$). 
The coordinate on $S^1$ is labelled by $x^3$ and we write $\hat{\lambda} = \bar{\lambda}_zdz + \lambda_{\bar{z}}d\bar{z}$ where $dz$ and $d\bar{z}$ are the dual one-forms to $\nabla_z$ and $\nabla_{\bar{z}}$ respectively. In a local orthonormal frame the metric tensor has components $g^{z\bar{z}} = 2, g_{z\bar{z}} = 1/2$. The two unit vectors $\partial_t$ and $\partial_n$ denote the tangent and outward normal to $\partial\Sigma$ respectively, so:
\begin{equation}
\hat{\lambda}_t = \hat{\lambda}(\partial_t),\quad 
\hat{\lambda}_n = \hat{\lambda}(\partial_n), \quad 
A_t = A(\partial_t),\quad
A_n = A(\partial_n).
\end{equation}

\paragraph{Vector multiplets.}
A twisted 3d $\mathcal{N}=2$ vector multiplet has fields $(A_\mu, \lambda,\bar{\lambda},\lambda_{\bar{z}},\bar{\lambda}_z, D)$. The supersymmetry transformation laws are:
\begin{equation}
\begin{aligned}
\delta A_3 &= -\frac{i}{2}(\epsilon\bar{\lambda} - \bar{\epsilon}\lambda),\\
\delta A_{\bar{z}} &= -\frac{i}{2}\bar{\epsilon}\lambda_{\bar{z}},\\
\delta A_z &= -\frac{i}{2}\epsilon\bar{\lambda}_z,\\
\delta \sigma &= \frac{1}{2}(\epsilon \bar{\lambda} - \bar{\epsilon}\lambda),
\end{aligned}\qquad
\begin{aligned}
\delta \lambda &= 2i\epsilon F_{z\bar{z}} - \epsilon \partial_3\sigma +\epsilon D,\\
\delta \bar{\lambda} &= 2i\bar{\epsilon} F_{z\bar{z}} + \bar{\epsilon} \partial_3\sigma + \bar{\epsilon} D,\\
\delta \bar{\lambda}_z &= -2i\bar{\epsilon} F_{z3} +2\bar{\epsilon} \partial_z\sigma,\\
\delta \lambda_{\bar{z}} &= -2i\epsilon F_{\bar{z}3} +2\epsilon \partial_{\bar{z}}\sigma,\\
\delta D &= \epsilon \nabla_{\bar{z}}\bar{\lambda}_z - \bar{\epsilon}\nabla_z\lambda_{\bar{z}} + \frac{\bar{\epsilon}}{2}\partial_3\lambda + \frac{\epsilon}{2}\partial_3\bar{\lambda}.
\end{aligned}
\end{equation}
The vector multiplet Lagrangian is
\begin{equation}
\mathcal{L}_{\text{vec}} =  \mathcal{L}_{\text{vec}}^{B} + \mathcal{L}_{\text{vec}}^{F},
\end{equation}
with the bosonic and fermionic parts given by
\begin{equation}\label{eq:lagrangian_vector}
\begin{aligned}
 \mathcal{L}_{\text{vec}}^{B} 
&=4|F_{z\bar{z}}|^2 + 4F_{z3}F_{\bar{z}3} + \nabla_\mu\sigma\nabla^\mu\sigma -D^2,\\
\mathcal{L}_{\text{vec}}^{F}
&=\frac{1}{2}(\bar{\lambda}\nabla_3\lambda +\lambda\nabla_3\bar{\lambda}+\lambda_{\bar{z}}\nabla_3\bar{\lambda}_z+\bar{\lambda}_z\nabla_3\lambda_{\bar{z}} \\
&\quad+2\lambda\nabla_{\bar{z}}\bar{\lambda}_z + 2\bar{\lambda}_z\nabla_{\bar{z}}\lambda-2\bar{\lambda}\nabla_z\lambda_{\bar{z}}-2\lambda_{\bar{z}}\nabla_z\bar{\lambda}).
\end{aligned}
\end{equation}
The vector multiplet Lagrangian can be written as a $Q$-exact term
\begin{equation}
\epsilon\bar{\epsilon} \mathcal{L}_{\text{vec}} = \delta_{\epsilon}\delta_{\bar{\epsilon}}(\lambda\bar{\lambda} -\lambda_{\bar{z}}\bar{\lambda}_z - 4\sigma D).
\end{equation}
We refer the reader to appendix \ref{ap:lagrangians} for more details on these Lagrangians. In particular, we have performed some integration by parts when compared to the usual twisted action in \textit{e.g.} \cite{benini2015topologically}, and the boundary terms must be shown to vanish.

We may also introduce a Fayet-Iliopoulos term:
\begin{equation}\label{eq:3d_FI}
    \mathcal{L}_{\text{FI}}^{3d} = \frac{1}{2\pi\text{Vol}(HS^2)} \left(-i A_3^{(T)} F_{12} + \zeta D \right),
\end{equation}
where $A_3^{(T)}$ and $\zeta$ are a flat connection and real mass (FI parameter) for a background vector multiplet coupling to the topological symmetry. The above action is manifestly gauge invariant and supersymmetric under the boundary conditions we introduce in the following. We also include a boundary FI term:
\begin{equation}\label{eq:2d_FI}
    \mathcal{L}_{\text{FI}}^{2d} = - \frac{i}{\text{Vol}(\partial HS^2)}(A_3+i\sigma)(A_3^{(T)} + i\zeta),
\end{equation}
which we will see is a constant for our choice of boundary condition.\footnote{This term is necessary to include in order to preserve gauge invariance and supersymmetry if one were to consider Neumann boundary conditions on the gauge field and make the background vector multiplet for the topological symmetry dynamical.} Nevertheless, we shall see it is useful to include, as it naturally complexifies the fugacity for the topological symmetry in the index we consider in the following.

\paragraph{Chiral multiplets.}
A chiral multiplet, once twisted, has fields $(X, \psi,\bar{\psi},\psi_{\bar{z}},\bar{\psi}_z, F_{\bar{z}})$. We will primarily consider only chiral multiplets with $U(1)_R$-charge zero in this work, for which the supersymmetry transformation laws are:\footnote{Where the vector multiplet fields act in the appropriate representation.}
\begin{equation}\label{eq:chiraltransformation}
\begin{aligned}
\delta X &= \bar{\epsilon}\psi,\\
\delta \bar{X} &= \epsilon\bar{\psi},\\
\delta \psi &= \epsilon(\nabla_3 + iA_3 - \sigma)X,\\
\delta \bar{\psi} &= \bar{\epsilon}(\nabla_3 - iA_3 + \sigma) \bar{X},
\end{aligned}\qquad
\begin{aligned}
\delta \bar{\psi}_z &= 
2\bar{\epsilon} (\nabla_z - iA_z)\bar{X} - \epsilon \bar{F}_z,\\
\delta \psi_{\bar{z}} &= -2\epsilon(\nabla_{\bar{z}} + iA_{\bar{z}})X + \bar{\epsilon} F_{\bar{z}},\\
\delta \bar{F}_z &= -\bar{\epsilon}(\nabla_3 - iA_3 + \sigma)\bar{\psi}_z + 2\bar{\epsilon}(\nabla_z - iA_z)\bar{\psi} - \bar{\epsilon}\bar{\lambda}_z \bar{X},\\
\delta F_{\bar{z}} &= \epsilon(\nabla_3 + iA_3 - \sigma)\psi_{\bar{z}} +2 \epsilon(\nabla_{\bar{z}} + iA_{\bar{z}})\psi +\epsilon\lambda_{\bar{z}}X.
\end{aligned}
\end{equation}
Note that
\begin{equation}
    [\delta_{\bar{\epsilon}}, \delta_{\epsilon}] = \epsilon\bar{\epsilon}( A_3 + i\sigma + \nabla_3),
\end{equation}
and that $A_3 + i\sigma$ may be regarded as a complexified gauge transformation parameter.

The chiral multiplet Lagrangian is
\begin{equation}\label{eq:chiral_lagrangian}
\mathcal{L}_{\text{chi}} = \mathcal{L}_{\text{chi}}^{B} + \mathcal{L}_{\text{chi}}^{F},
\end{equation}
with the bosonic and fermionic parts given by:
\begin{equation}
\begin{aligned}
\mathcal{L}_{\text{chi}}^{B}=&
(\nabla_3 - iA_3)\bar{X}(\nabla_3 + iA_3)X + 4(\nabla_z - iA_z)\bar{X}(\nabla_{\bar{z}} + iA_{\bar{z}})X \\
&+ \sigma^2|X|^2 + D|X|^2 +2iF_{z\bar{z}}|X|^2 - F_{\bar{z}}\bar{F}_z,\\
\mathcal{L}_{\text{chi}}^{F} =&
-\bar{\psi}(\nabla_3 + iA_3 + \sigma)\psi - \bar{\psi}_z(\nabla_3 + iA_3 - \sigma)\psi_{\bar{z}} + 2\bar{\psi}(\nabla_z + iA_z)\psi_{\bar{z}}  \\ &-2\bar{\psi}_z(\nabla_{\bar{z} } + iA_{\bar{z}})\psi +\bar{\lambda}_{z}\bar{X}\psi_{\bar{z}} +\bar{\psi}_z\lambda_{\bar{z}}X + \bar{\psi}\lambda X - \bar{X}\bar{\lambda}\psi.
\end{aligned}
\end{equation}
The chiral multiplet Lagrangian may be written in the $Q$-exact form:
\begin{equation}
\begin{aligned}
\epsilon\bar{\epsilon}\mathcal{L}_{\text{chi}} &= \delta_{\epsilon}\delta_{\bar{\epsilon}}(-\bar{\psi}\psi +\bar{\psi}_z\psi_{\bar{z}}+2\bar{X}\sigma X).
\end{aligned}
\end{equation}
For completeness, we note that in the $C$-twisted 3d $\mathcal{N}=4$ Lagrangian, there is also a $G$ adjoint $\mathcal{N}=2$ chiral multiplet in the $\mathcal{N}=4$ vector multiplet. For the $C$-twist with $U(1)_R = U(1)_C$ it has $R$ charge $2$, and thus has components: $(\phi_{\bar{z}}, \bar{\phi}_z, \eta,\bar{\eta},\eta_{\bar{z}},\bar{\eta}_z, F)$. The adjoint Lagrangian is
\begin{equation}
\mathcal{L}_{\text{adj}} = 4\nabla_z \phi_{\bar{z}}\nabla_{\bar{z}}\bar{\phi}_z  + \nabla_3\phi_z\nabla_3\phi_{\bar{z}} +\bar{\eta}\nabla_3\eta + \eta_{\bar{z}}\nabla_3\bar{\eta}_z+2\eta \nabla_{\bar{z}} \bar{\eta}_z -2\bar{\eta}\nabla_z\eta_{\bar{z}} - F\bar{F} \,.
\end{equation}

\paragraph{Global symmetries.} Throughout this work, we introduce additional background vector multiplets in order to grade by flavour symmetries.\footnote{Alternatively, we could also introduce twisted periodicities for fields around $S^1$, see \cite{Hori:2014tda} for further details.} The supersymmetry transformation is thus modified by replacing $A_3\to A_3 + A_3^{(B)}$ and $\sigma\to \sigma + m$ where $A_3^{(B)}, m$ are the vev of the background gauge field and scalar.

\subsection{\texorpdfstring{$\mathcal{N}=4$}{} supersymmetry} 

In this work we will be concerned primarily with 3d $\mathcal{N}=4$ supersymmetric gauge theories specified by a compact connected gauge group $G=U(1)^r$ and a quaternionic linear representation $Q = T^{*}R$. Such theories admit a flavour symmetry $G_H \times G_C$ where:
\begin{itemize}
    \item $G_C$ is the topological symmetry acting on monopole operators. It has a maximal torus $T_C$, which is coupled via an FI-term in the UV, which can be enhanced in the IR to a non-Abelian symmetry $G_C$. Concretely:
    \begin{equation}
        T_C = \text{Hom}(\pi_1(G),U(1)).
    \end{equation}
    \item $G_H$ is the Higgs branch flavour symmetry acting on the hypermultiplets, and is equal to
    \begin{equation}
        G_H = N_{U(R)}G / G,
    \end{equation}
    where $N_{U(R)}$ is the normaliser of the unitary representation $U(R)$. We denote the maximal torus of $G_H$ by $T_H$. 
\end{itemize}

The $R$-symmetry is $SU(2)_H \times SU(2)_C$ the Cartan subgroup of which we denote by $U(1)_H\times U(1)_C$. We choose a convention where fields have integer charge under these symmetries. We will often be interested in turning on a fugacity for the diagonal combination $T_t = U(1)_H - U(1)_C$, which breaks the supersymmetry to $\mathcal{N}=2$. 

\paragraph{Example: $\text{SQED}[N]$.}
We will consider supersymmetric QED with $N$ flavours as a running example throughout this work. This theory is specified by a gauge group $G = U(1)$ and the representation $R = \mathbb{C}^{N}$. We will denote the scalars in the hypermultiplets by $(X,Y)$. The flavour symmetries are then $G_C = U(1)$ and $G_H = PSU(N)$. We introduce real mass parameters $(m_1,\ldots,m_N)$ with $\sum m_i = 0$ for $G_H$, a real FI parameter $\zeta$, and a real mass $m_t$ for $T_t$. In our conventions, the hypermultiplet scalars $(X,Y)$ have charge $+1$ under $U(1)_H$, and so $+1/2$ under $T_t$.

\paragraph{Topological twist.}

3d $\mathcal{N}=4$ theories admit different choices of topological twist, corresponding to a $U(1)_R$ $R$-symmetry of an $\mathcal{N}=2$ subalgebra, which we use to twist. For example, choosing $U(1)_R$ to be $U(1)_H$ or $U(1)_C$ results in two distinct (fully topological) twists, related by 3d mirror symmetry \cite{intriligator1996mirror}. In this work, we twist with respect to $U(1)_C$ along $HS^2$. This breaks the $R$-symmetry down to $SU(2)_H\times U(1)_C$, which our boundary conditions further break to $U(1)_H \times U(1)_C$. The $C$-twist preserves four supercharges on a manifold without boundary; our boundary condition breaks half of the twisted supercharges. We refer to section §$6.1$ of \cite{Closset_2016} for further details. We note that the hypermultiplet scalars have zero $U(1)_C$ $R$-charge, so that the supersymmetry transformations in section \ref{eq:susytransforms} hold.

\paragraph{Field content \& Lagrangian.}

The off-shell\footnote{The full 3d $\mathcal{N}=4$ supersymmetry algebra cannot be closed off-shell by adding finitely many auxiliary fields. We have added auxiliary fields to close a 3d $\mathcal{N}=2$ subalgebra of the 3d $\mathcal{N}=4$ algebra.} field content of a 3d $\mathcal{N}=4$ theory consists of an $\mathcal{N}=4$ vector multiplet for $G$, which may be split as follows
\begin{itemize}
    \item $\mathcal{N}=2$ vector multiplet $(A_\mu, \lambda,\bar{\lambda},\lambda_{\bar{z}},\bar{\lambda}_z,\sigma, D)$.
    \item $\mathcal{N}=2$ adjoint chiral multiplet $(\phi_{\bar{z}}, \bar{\phi}_z, \eta,\bar{\eta},\eta_{\bar{z}},\bar{\eta}_z, F,\bar{F})$.
\end{itemize}
In addition, the hypermultiplets may be split into:
\begin{itemize}
\item $\mathcal{N}=2$ chirals $(X,\bar{X},\psi^{X},\bar{\psi}^{X}, \psi_{\bar{z}}^{X}, \bar{\psi}_z^{X}, F^{X}_{\bar{z}}, \bar{F}_z^{X})$.
\item $\mathcal{N}=2$ chirals $(Y,\bar{Y}, \psi^{Y},\bar{\psi}^{Y}, \psi_{\bar{z}}^{Y}, \bar{\psi}_z^{Y},F^{Y}_{\bar{z}}, \bar{F}_z^{Y})$.
\end{itemize}
where  the complex scalars $(X, Y)$ transform in the representation $T^*R = R \oplus R^*$ of $G$.

The Lagrangian can be schematically written as:
\begin{equation}
\mathcal{L} = \mathcal{L}_{\text{vec}} + \mathcal{L}_{\text{adj}} + \mathcal{L}_{X} + \mathcal{L}_{Y} + \mathcal{L}_{\text{Yukawa}}.
\end{equation}
The Yukawa term $\mathcal{L}_{\text{Yukawa}}$ descends from the superpotential $\phi \cdot \mu_{\mathbb{C}}$, where $\mu_{\mathbb{C}}$ is the complex moment map for the $G$ action on the hypermultiplet representation $T^*R$ (we return to this momentarily).  After integrating out the auxiliary fields, the potential term for the matter multiplets is given by:
\begin{equation}\label{eq:potential}
(\sigma^2 + |\phi_{\bar{z}}|^2)(\bar{X}\cdot X + \bar{Y}\cdot Y) + \frac{1}{4}(\bar{X}\cdot X - \bar{Y}\cdot Y)^2 + (X\cdot Y)(\bar{X}\cdot \bar{Y}).
\end{equation}

\subsubsection{Higgs branch geometry}\label{subsec:higgsbranch}
In this section we provide a minimal summary of the Higgs branch geometry of 3d $\mathcal{N}=4$ supersymmetric gauge theories. We denote the real and complex moment maps of the gauge symmetry $G$ by
\begin{equation}
    \mu_{\mathbb{R}}: \,T^*R \to \mathfrak{g}^*\,, \quad \mu_{\mathbb{C}}:\, T^*R \to \mathfrak{g}^*_{\mathbb{C}} \,.
\end{equation}
Note that the sum of the last two terms in the scalar potential \eqref{eq:potential} is precisely $|\mu_{\mathbb{R}}|^2 + |\mu_{\mathbb{C}}|^2$, and is $SU(2)_H$-invariant as $(\mu_{\mathbb{R}},\Re \mu_{\mathbb{C}},\Im \mu_{\mathbb{C}})$ transforms as a triplet under $SU(2)_H$. 

The Higgs branch $\mathcal{M}_H$ is then given by the hyper-K\"ahler quotient
\begin{equation}
    \mathcal{M}_H = \mu_{\mathbb{C}}^{-1}(0) \cap \mu_{\mathbb{R}}^{-1}(\zeta) / G \,,
\end{equation}
where throughout we assume it is possible to choose $\zeta$ in a fixed chamber $\mathfrak{C}_C$ such that $\mathcal{M}_H$ is fully resolved. Trading the real moment map condition for a stability condition (dependent on the choice of $\mathfrak{C}_C$) and a complexified gauge group action we may obtain an algebraic description of the Higgs branch as a symplectic variety\footnote{The `s' subscript denotes the stability condition which depends on the chamber $\mathfrak{C}_C$ containing $\zeta$.}
\begin{equation}
    \mathcal{M}_H = \mu^{-1}_{\mathbb{C}}(0)^{\text{s}} / G_{\mathbb{C}} \,.
\end{equation}
The coordinate ring of this variety is given by $\mathbb{C}[\mathcal{M}_H] = \left(\mathbb{C}[T^*R] / \langle\mu_{\mathbb{C}}\rangle\right)^{G_{\mathbb{C}}}$, we work with examples where this ring coincides with the cohomology of the Rozansky-Witten supercharge $Q_B$ (see e.g. \cite{dimofte2020mirror} for a discussion of this subtlety). $\mathcal{M}_H$ admits symplectic $G_H$ and $T_t$ actions with corresponding flavour symmetry real moment maps $\mu_{H,\mathbb{R}}$ and $\mu_{t,\mathbb{R}}$. 

We work with theories that admit a generic one-parameter subgroup of $T_H$, specified by real masses $m$ in $\mathfrak{t}_H$ lying in a fixed chamber $\mathfrak{C}_H$,\footnote{There are co-dimension $1$ loci in the space $\mathfrak{t}_H$ of mass parameters $m$ where the fixed loci in $\mathcal{M}_H$ of the one-parameter subgroup generated by $m$ is no longer isolated. A choice of chamber $\mathfrak{C}_H$ is a choice of connected component in the complement of these loci.} that gives rise to a finite set of isolated massive vacua/fixed points on the Higgs branch, where the gauge group is completely broken. We denote the set of vacua by $\mathcal{M}_H^{T_H} = \{ \alpha \}$.

The tangent bundle $T\mathcal{M}_H$ is given by the cohomology of the complex
\begin{equation}\label{eq:tangentbundle}
    0 \to \mathfrak{g}_{\mathbb{C}} \to T^*R \to \mathfrak{g}_{\mathbb{C}}^{\star} \to 0. 
\end{equation}
The first map is an infinitesimal complex gauge transformation and the second is the differential of the complex moment map. The complex thus encodes fluctuations of the hypermultiplets satisfying the moment map condition modulo gauge transformations. Writing $T=T_H\times T_t$ and taking the character of this complex we find:
\begin{equation}\label{eq:tangentcomplex}
    \text{ch}_T T\mathcal{M}_H = \text{ch}_T T^*R - \text{ch}_{T} \mu_{\mathbb{C}} - \text{ch}_T \mathfrak{g}_{\mathbb{C}}.
\end{equation}

The terms in the complex transform as weights of $G \times T_H \times T_t$. Introducing formal grading parameters $(s,x,t)$ for these symmetries, under our assumptions the vacua $\alpha$ ($T_H$ fixed points on $\mathcal{M}_H$) are labelled by $r= \text{rk}(G)$ weights $\rho_i = (\rho_{G,i}, \rho_{H,i},\rho_{t,i})$. These are the charges of the hypermultiplet scalars which acquire vevs in the vacuum $\alpha$. Solving the equations
\begin{equation}\label{eq:fugacity_subsitution}
    s^{\rho_{G,i}} x^{\rho_{H,i}} t^{\rho_{t,i}} = 1,
\end{equation}
for each $i=1,\ldots,r$ gives a unique solution $s \in \mathfrak{g}$ for each vacuum $\alpha$ which we denote by $s=s_{\alpha}$.\footnote{We note that, geometrically, this is equivalent to the action of the push-forward to the fixed point $\alpha$ under the Chern map of the vector bundles in equation \eqref{eq:tangentcomplex}.}
We denote the $T_{H} \times T_t$ character of the tangent bundle fibre at a fixed point thus obtained by evaluating \eqref{eq:tangentcomplex} at $s_{\alpha}$ by $T_{\alpha} \mathcal{M}_H$.

\paragraph{Example: \texorpdfstring{$\text{SQED}[N]$}{}.}
Working equivariantly with respect to $G \times T_H\times T_t$, which acts as $X_i\to s x_it^{1/2} X_i,\, Y_i\to s^{-1}x_i^{-1}t^{1/2} Y_i$, the character of the tangent bundle of $\mathcal{M}_H$ \eqref{eq:tangentcomplex} for $\text{SQED}[N]$ may be expressed as: 
\begin{equation}\label{eq:sqedtangent}
    T\mathcal{M}_H = t^{1/2}s(x_1 + \ldots + x_N) + t^{1/2} s^{-1} (x_1^{-1} + \ldots + x_N^{-1}) - (1+t).
\end{equation}
In the chamber $\mathfrak{C}_C$ where $\zeta > 0$, the massive vacuum $\alpha$ corresponds to non-vanishing vev $|X_\alpha|^2 = 2\zeta$. We retain this choice of chamber throughout this work. Thus, we solve the equation $s x_{\alpha} t^{\frac{1}{2}}=1$ to find $s=x_{\alpha}^{-1}t^{-\frac{1}{2}}$ which then gives the expression
\begin{equation}
    T_{\alpha} \mathcal{M}_H = \sum_{i \neq \alpha} \frac{x_i}{x_{\alpha}} + t \frac{x_{\alpha}}{x_i},
\end{equation}
for the character of the fibre of the tangent bundle at a fixed point.

\section{Hemisphere index}\label{sec:twistedtheories}

In this section, we discuss quasi-topologically twisted theories on $HS^2 \times S^1$. We describe the boundary conditions of interest; the exceptional Dirichlet (or thimble) boundary conditions and introduce the central object of our study: the twisted hemisphere index.

\subsection{Hemisphere geometry}\label{sec:hem_geom}

We work on the background $HS^2\times S^1$, with product metric:
\begin{equation}
    ds^2 = d\theta^2 + \sin^2 \theta \,d\phi^2 + dx^3dx^3,
\end{equation}
where:
\begin{equation}
x^3 \in [0,\beta], \quad \theta\in [0,\pi/2], \quad \phi\in[0,2\pi].
\end{equation}
We will often conformally map $HS^2$ to the unit disc $|w|\leq 1$ on the complex plane and pull back the coordinate $w\in \mathbb{C}$ to obtain a global coordinate on $HS^2$. On fermions, we impose Ramond (periodic) boundary conditions around $\partial HS^2 = S^1$ as usual for twisted theories.

\paragraph{Boundary conditions.}

We first describe boundary conditions for 3d $\mathcal{N}=2$ theories preserving the two twisted supercharges. We impose a Dirichlet boundary condition for the vector multiplet:
\begin{equation}\label{eq:dirichletvector}
A_3 + i\sigma = \text{const}, \quad A_t = 0,  \quad \lambda_t = 0, \quad \lambda - \bar{\lambda}=0.
\end{equation}
This boundary condition trivialises the  $U(1)^r$ gauge bundle in a neighbourhood of the boundary torus. We note also:

\begin{itemize}
    \item The constant boundary value of $A_3+i\sigma$ is arbitrary for a pure gauge theory, but in a theory with matter it must take certain values in order to preserve flavour symmetries at the boundary and be compatible with localisation. We return to this momentarily.
    \item The topological class of the trivialisation is labelled by the winding numbers (with respect to a fixed trivialisation) around the $\partial (HS^2 \times S^1) = T^2$. Different winding numbers around the $x^3$-cycle are gauge equivalent configurations. In contrast, the winding number of the trivialisation around $\partial HS^2 = S^1$ leads to physically distinct configurations labelled by the monopole flux across the hemisphere:
    \begin{equation}
    k = -\frac{1}{2\pi}\int_{HS^2} F_{12} = \frac{i}{\pi}\int_{HS^2} F_{z\bar{z}} \in \mathbb{Z}^r.
    \end{equation}
    The index we compute will involve a sum over the monopole sectors $k$.
    \item In order to properly define the angular momenta of the fields, one needs to lift $SO(2)_E$ rotations of the disc to the total space of the $G$-bundle over $HS^2$. We choose a lift compatible with the Dirichlet boundary condition for the gauge field, preserving the trivialisation on the boundary torus. \textit{E.g.} for $G=U(1)$, in the flux $k$ sector, the fibre of the associated charge $+1$ line bundle above the origin carries a weight $k$ representation of $SO(2)_E$.
\end{itemize}

For the chiral multiplets, the two basic boundary conditions we impose are: 
\begin{equation}\label{eq:chiral_boundary_conditions}
\begin{aligned}
\text{Dirichlet:}& \quad X = c,\,
\quad  \psi = \bar{\psi} = 0.\\
\text{Neumann:}& \quad (\nabla_{\bar{z}} + iA_{\bar{z}})X = (\nabla_z - iA_z)\bar{X} = 0,\\
& \quad (\nabla_{\bar{z}} + iA_{\bar{z}})\psi + \frac{1}{2}\lambda_{\bar{z}}X = (\nabla_z - iA_z)\bar{\psi} -\frac{1}{2}\bar{\lambda}_z\bar{X} =  0,\\
& \quad \psi_{\bar{z}} = \bar{\psi}_z = F_{\bar{z}}  = \bar{F}_z = 0,
\end{aligned}
\end{equation}
where the vector multiplet fields act in the appropriate representation, and $c$ is a constant.
For later reference, we will also be interested in imposing Dirichlet boundary conditions for the adjoint chiral in the 3d $\mathcal{N}=4$ vector multiplet:
\begin{equation}
\phi_{\bar{z}} = \bar{\phi}_z = \eta_{\bar{z}} = \bar{\eta}_z = 0.
\end{equation}

Note that these are \textit{not} the analogue of the $A$-brane boundary conditions in 2d (one obtains an $A$-twisted $(2,2)$ theory after reducing along the $S^1$), which would \textit{e.g.} impose Dirichlet on $\text{Re}(X)$ and Neumann on $\text{Im}(X)$, or vice versa, forcing the matter fields to lie in a \textit{real} Lagrangian. The boundary conditions on the complex scalar component $X$ is the same as that of the $B$-brane type boundary conditions preserving $(0,2)$ supersymmetry that are imposed on the rigid supersymmetry background for the half superconformal index \cite{bullimore2021boundaries}. However, the fermionic boundary conditions are different. In fact, on flat space, they are `sick' in the sense of appendix B.2 of \cite{dedushenko2021interfaces} (see also \cite{dedushenko2020gluing,dedushenko2021gluing}). That is, there are an infinite number of fermionic zero modes, which in flat space would lead to a vanishing partition function. On the twisted $S^1 \times HS^2$ background, they correspond to the infinite number of holomorphic functions on the disk (see subsection \ref{sec:localisation}) and would also naively contribute $0^\infty$ to the fermionic path integral. However, we shall see that the definition of the index requires that the fields obey a twisted periodicity around $S^1$, equivalent to switching on an Omega background, see \textit{e.g.} \cite{benini2014higgs}. This lifts the zero modes, weighting them by their angular momenta. Thus, these boundary conditions are consistent in the twisted background for the theory on $HS^2 \times S^1$. Note that similar boundary conditions were considered in \cite{Longhi:2019hdh}, for 4d theories on $D^2 \times T^2$ where $D^2$ is the disk and $T^2$ the torus, and in \cite{pittelli2020localisation} for 3d theories on $H^2 \times S^1$, where $H^2$ is the hyperbolic plane. In both situations the naive sickness of the boundary conditions is also ameliorated by the presence of an effective Omega background.\footnote{We thank Mykola Dedushenko for clarifying these issues to us.}

There are some subtleties arising from the boundary value of $c$ for the Dirichlet boundary condition:
\begin{itemize}
    \item If we assign $c=0$ for all chiral multiplets with a Dirichlet boundary condition (or if the theory is a pure gauge theory), there is an additional boundary flavour symmetry $G_{\partial}$ arising from the bulk gauge symmetry. The boundary value of $A_3+i\sigma$ may be regarded as a background for this symmetry, and set arbitrarily. We shall see this phenomenon in the Coulomb branch localisation scheme.
    \item If $c\neq 0$, in order to preserve the supersymmetric boundary condition $\psi = \bar{\psi}=0$, from \eqref{eq:chiraltransformation} we see that in the absence of background fields for flavour symmetries, one must take $A_3 = \sigma  = 0$. If one has turned on background vector fields for flavour symmetries under which $X$ is charged, then $A_3$ and $\sigma$ must cancel them. For example, suppose $X$ has charges $(1,q_f)$ under a $G_{\partial}$ and a flavour symmetry $F$. If we have introduced a background holonomy $A_3^f$ and mass $m$, then we must have: $A_3 = -q_f A_3^f$ and $\sigma =-q_f m$. This ensures a combination of $G_{\partial}$ and $F$ is preserved, generated by $J_{\partial}-q_f J_F$, which we take to be a re-definition of $F$. In our work, we will ultimately be interested in turning on a non-zero value of $c$ for the fields which take on vevs in a Higgs vacua. We shall see the above is compatible with the Higgs branch localisation scheme.
\end{itemize}

\paragraph{$\mathcal{N}=(2,2)$ boundary conditions.}

In this work, we will primarily be concerned with $\mathcal{N}=(2,2)$ boundary conditions for $\mathcal{N}=4$ theories. These were first studied systematically in \cite{bullimore2016boundaries}, to which we refer the reader for more details. 

\begin{itemize}
    \item On the 3d $\mathcal{N}=4$ vector multiplet, we impose the $\mathcal{N}=(2,2)$ completion of the Dirichlet boundary condition for the vector multiplet. This consists of the Dirichlet boundary condition \eqref{eq:dirichletvector} for the $\mathcal{N}=2$ vector multiplet, and a Dirichlet boundary condition for the the adjoint $\mathcal{N}=2$ chiral multiplet, which is set to zero at the boundary.
    \item On the 3d $\mathcal{N}=4$ hypermultiplets, we impose Neumann-Dirichlet type boundary conditions. These are specified by a holomorphic Lagrangian splitting (polarisation) of the representation $Q=T^*R$. Writing $R=\mathbb{C}^N$, this is specified by a sign vector $\varepsilon \in \{\pm\}^N$ such that, for $i=1,\ldots, N$:
    \begin{equation}\label{eq:polarisation}
        (X_{\varepsilon_i}, Y_{\varepsilon_i}) =
        \begin{cases}
        (X_i,Y_i) \quad &\text{if } \varepsilon_i = + \\
        (Y_i, -X_i)\quad &\text{if } \varepsilon_i = - \\
        \end{cases}.
    \end{equation}
    The boundary condition imposes Neumann boundary conditions for the $\mathcal{N}=2$ chirals with scalar components $X_{\varepsilon_i}$, and Dirichlet for those with scalar components $Y_{\varepsilon_i}$. For the $Y_{\varepsilon_i}$, we must also specify their boundary values.
\end{itemize}

\subsection{Exceptional Dirichlet boundary conditions}\label{subsec:ed}
We now turn to a distinguished class of boundary conditions for $\mathcal{N}=4$ theories known as exceptional Dirichlet. This set of boundary conditions is naturally associated to Lagrangian submanifolds of the Higgs branch geometry of subsection \ref{subsec:higgsbranch} as we now review (we refer the reader to \cite{bullimore20223d, bullimore2021boundaries, bullimore2016boundaries, Crew:2021ipc} for further details). The boundary conditions are designed to mimic the presence of an isolated vacuum $\alpha$ at infinity, at least for quantities amenable to supersymmetric localisation.

The exceptional Dirichlet boundary condition, denoted $D_{\alpha}$, is given by a Dirichlet boundary condition on the vector multiplet, and Neumann-Dirichlet boundary condition on the hypermultiplets. It is expected that the set of UV boundary conditions described here flow to thimble boundary conditions in the IR Rozansky-Witten $\sigma$-model on $\mathcal{M}_H$. Given a set of mass parameters $m$ in a fixed chamber $\mathfrak{C}_H \subset \mathfrak{g}_H^{*}$ we obtain a symplectic $\mathbb{C}^{\times}_m$ action on $\mathcal{M}_H$ with an associated Morse function $h = m \cdot \mu_{H,\mathbb{R}}$ with critical points at the isolated vacua $\{\alpha\}$. We obtain a set of distinguished Lagrangian subvarieties $\mathcal{L}_{\alpha} \subset \mathcal{M}_H$ from Morse flow of $h$ or, equivalently, defined by the closures of attracting sets to fixed points
\begin{equation}
    \text{Attr}_{\alpha} = \{ p \in \mathcal{M}_H \, : \, \lim_{t \to 0} m(t) \cdot p = \alpha \} \,,
\end{equation}
and setting $\mathcal{L}_\alpha = \overline{\text{Attr}_{\alpha}}$. Accordingly the tangent weights in $T_{\alpha} \mathcal{M}_H$ split into positive and negative $T_H$ weights so that
\begin{equation}
    T_{\alpha}\mathcal{M}_H = T_{\alpha}^+\mathcal{M}_H + T_{\alpha}^-\mathcal{M}_H \,,
\end{equation}
with $T_{\alpha} \mathcal{L}_{\alpha} = T_{\alpha}^+\mathcal{M}_H$. 

The chamber dependent Neumann-Dirichlet boundary condition on the hypermultiplets, specified by $T^*R$, corresponds to a decomposition of weights
\begin{equation}
    T^*R = Q^- + Q^0 + \bar{Q}^0 + Q^+ \,,
\end{equation}
where, after the fugacity substitution \eqref{eq:fugacity_subsitution}, we recover the positive; zero and symplectic conjugate;\footnote{Here, symplectic conjugate refers to the pairing of weights $(\omega,t \omega^{-1})$ via the holomorphic symplectic form.} and negative weights in the tangent bundle respectively. In particular, the chiral multiplets corresponding to weights in $Q^-$ and $Q^0$ are assigned Dirichlet boundary conditions, and those in $Q^+$ and $\bar{Q}^0$ Neumann. This determines a polarisation \eqref{eq:polarisation}. Further, the chirals in $Q^-$ are set to zero on the boundary, whilst those in $Q^0$ are set to a constant non-zero value, for example their values in the vacuum $\alpha$. In turn, the boundary value of $A_3+i\sigma$ in the vector multiplet must be tuned to cancel any connection and mass for the flavour symmetries coupled to those chirals in $Q^0$.

\subsubsection{\texorpdfstring{$\text{SQED}[N]$}{} example}\label{sec:sqed_ed}
We return to our illustrative example of supersymmetric QED with $N$ flavours. We have $\alpha = 1, \ldots, N$ vacua. The $2N$ weights of $T^{*}R$ may be expressed as a character\footnote{We conflate notation for characters and weight spaces throughout this work.}
\begin{equation}
    Q = T^*R = t^{1/2}s(x_1 + x_2 + \ldots + x_N) + t^{1/2} s^{-1}(x_1^{-1} + \ldots + x_N^{-1}) .
\end{equation}
Recall the Chern root evaluation \eqref{eq:fugacity_subsitution} is given by $s=x_{\alpha}^{-1}t^{-\frac{1}{2}}$. We choose the chamber:
\begin{equation}
    \mathfrak{C}_H = \{m_1 < m_2 < \ldots < m_N\},
\end{equation}
so that for each $\alpha$ the corresponding split into $T_H$ weight spaces is
\begin{equation}
\begin{split}
    &Q^{-} = t^{1/2} s(x_1 + \ldots + x_{\alpha-1}) +  t^{1/2} s^{-1}(x_{\alpha+1}^{-1} + \ldots + x_{N}^{-1}) , \\
    &Q^{0} = t^{1/2} s x_{\alpha} , \quad \bar{Q}^{0} = t^{1/2} s^{-1} x_{\alpha}^{-1}, \\
    &Q^{+} = t^{1/2}s(x_{\alpha+1} + \ldots + x_{N}) + t^{1/2} s^{-1}(x_1^{-1} + \ldots + x_{\alpha-1}^{-1}) .
\end{split}    
\end{equation}
After evaluating at a fixed point, we have the corresponding splitting of the tangent bundle
\begin{equation}
    T_{\alpha}\mathcal{M}_H = T^{+}_{\alpha}\mathcal{M}_H + T^{-}_{\alpha}\mathcal{M}_H ,
\end{equation}
where 
\begin{equation}
    T^{+}_{\alpha}\mathcal{M}_H = \sum_{i>\alpha} \frac{x_i}{x_{\alpha}}+t \sum_{i<\alpha} \frac{x_{\alpha}}{x_i} \,, \quad T^{-}_{\alpha}\mathcal{M}_H = \sum_{i < \alpha} \frac{x_i}{x_{\alpha}} + t \sum_{i> \alpha} \frac{x_{\alpha}}{x_i}.
\end{equation}
Geometrically, the Lagrangians $\mathcal{L}_{\alpha} \subset \mathcal{M}_H = T^* \mathbb{P}^{N-1}$ correspond to the conormal bundles of the Schubert cells.

We now implement the boundary conditions $D_{\alpha}$ in our hemisphere setup for this example. The Neumann-Dirichlet boundary condition on the hypermultiplets is specified by a polarisation vector $\varepsilon = \{-\ldots-+\ldots +\}$, where the first $\alpha$ components are `$-$'. For the $\mathcal{N}=2$ chirals assigned Dirichlet boundary conditions, namely $X_1, \ldots X_{\alpha}, Y_{\alpha+1},\ldots Y_N$, only $X_{\alpha}$ is assigned a non-zero boundary value $c$.

For the vector multiplet, we impose the $(2,2)$ Dirichlet boundary condition described in the previous section. Note that we must choose:
\begin{equation}\label{eq:substitution_fugacity}
    A_3+A_3^{\alpha}+A_3^t/2 = 0, \qquad\sigma + m_\alpha + m_t/2 = 0,
\end{equation}
in order to preserve supersymmetry since we have set $X_{\alpha}=c$. This boundary condition then preserves the flavour symmetry $T_C\times T_H \times T_t$, where as above $T_H$ is redefined by the generator of $G_{\partial}$ (see also §3.3 of \cite{bullimore2021boundaries}).

\subsection{Hemisphere index}
For a 3d $\mathcal{N}=4$ theory, the twisted hemisphere index is given by:
\begin{equation}\label{eq:index}
\mathcal{Z} =  \Tr_{\mathcal{H}}(-1)^F e^{-\beta H}q^J t^{\frac{R_H-R_C}{2}} x_i^{F_i} \xi^{F_C}
\end{equation}
which is a trace over the Hilbert space on $HS^2$, graded by fermion number and global symmetries. As usual, this can be computed as path integral on the topologically twisted background outlined above on $HS^2\times S^1$, where the fugacities are implemented via introducing background holonomies and masses. More precisely:
\begin{equation}
x_i = e^{i \beta (A_3^i +i m_i)},\quad  t = e^{i\beta (A_3^t+im_t)},\quad  \xi = e^{i\beta (A_3^{(T)} +i \zeta)}.
\end{equation}
We set also $q=e^{\ep}$, and implement the twisted periodicity due to the angular momentum via imposing in the path integral that fields must obey:
\begin{equation}\label{eq:twistedperiodicity}
   \Phi(x^3 = \beta) = q^{-J} \Phi(x^3 = 0). 
\end{equation}
The index preserves both twisted supercharges. We will be interested in the set of indices for a given theory for each of its exceptional Dirichlet boundary conditions $D_{\alpha}$, which we denote $\mathcal{Z}_{D_{\alpha}}$. For an $\mathcal{N}=2$ theory, the index is defined similarly.

We will compute this index using both Coulomb and Higgs branch localisation techniques. In the Coulomb branch localisation, we will first compute the index for $c=0$ in the exceptional Dirichlet boundary condition: \textit{i.e.} the boundary vevs of $\mathcal{N}=2$ chirals with Dirichlet boundary conditions are set to $0$. In this case, the index will also be graded with a fugacity
\begin{equation}
    s=e^{i\beta(A_3+i\sigma)|_{\partial}}
\end{equation} 
for the boundary $G_{\partial}$ symmetry, where $A_3+i\sigma|_{\partial}$ is a shorthand for the boundary values. Turning on $c \neq 0$, recall the sum of all flavour connections coupled to the chiral must sum to zero in order to preserve supersymmetry (\textit{e.g.} \eqref{eq:substitution_fugacity} for SQED[$N$]). In terms of fugacities, in the index we perform a substitution for $s$ in terms of $x_i$ and $t$, which mirrors the substitution \eqref{eq:fugacity_subsitution} in the tangent space character. It also neatly accounts for the redefinition of symmetries, as described in section \ref{sec:hem_geom} and in  §3.3 of \cite{bullimore2021boundaries}.

\subsubsection{Relation with half superconformal index}\label{subsubsubsec:halfindexrelation}

In the following we will see that the expression we obtain for the twisted hemisphere index \eqref{eq:index} is functionally identical to the formula for the half superconformal index on $HS^2\times S^1$:
\begin{equation}
    \mathcal{Z}_{\text{SC}} = \text{Tr}_{\mathcal{H}_{\text{SC}}} q^{J+ \frac{R_H+R_C}{4}} t^{\frac{R_H-R_C}{2}} x_i^{F_i} \xi^{F_C},
\end{equation}
provided the same boundary conditions are imposed. $\mathcal{Z}_{\text{SC}}$ is computed as a path integral with the same rigid supergravity background as the superconformal index \cite{kim2009complete, imamura2011index}, and $\mathcal{H}_{\text{SC}}$ denotes the Hilbert space on the hemisphere in this background. Notably, the fermions have Neveu-Schwarz (anti-periodic) boundary conditions around the spatial boundary $\partial HS^2$. The hemisphere partition function is related by an exponential prefactor (which we will also derive) to a count of boundary local operators in the cohomology of the same supercharge as for the usual superconformal index, originally studied in \cite{gadde2014walls, gadde2016fivebranes, dimofte2018dual}.

\begin{figure}[ht!]
    \centering

\tikzset{every picture/.style={line width=0.75pt}} 

\begin{tikzpicture}[x=0.7pt,y=0.7pt,yscale=-1,xscale=1]

\draw  [draw opacity=0][dash pattern={on 0.84pt off 2.51pt}] (510,160) .. controls (510,160) and (510,160) .. (510,160) .. controls (498.95,160) and (490,142.09) .. (490,120) .. controls (490,97.91) and (498.95,80) .. (510,80) -- (510,120) -- cycle ; \draw  [color={rgb, 255:red, 208; green, 2; blue, 27 }  ,draw opacity=1 ][dash pattern={on 0.84pt off 2.51pt}] (510,160) .. controls (510,160) and (510,160) .. (510,160) .. controls (498.95,160) and (490,142.09) .. (490,120) .. controls (490,97.91) and (498.95,80) .. (510,80) ;  
\draw  [draw opacity=0] (510,80) .. controls (510,80) and (510,80) .. (510,80) .. controls (510,80) and (510,80) .. (510,80) .. controls (521.05,80) and (530,97.91) .. (530,120) .. controls (530,142.09) and (521.05,160) .. (510,160) -- (510,120) -- cycle ; \draw  [color={rgb, 255:red, 208; green, 2; blue, 27 }  ,draw opacity=1 ] (510,80) .. controls (510,80) and (510,80) .. (510,80) .. controls (510,80) and (510,80) .. (510,80) .. controls (521.05,80) and (530,97.91) .. (530,120) .. controls (530,142.09) and (521.05,160) .. (510,160) ;  
\draw  [draw opacity=0] (236,160) .. controls (236,160) and (236,160) .. (236,160) .. controls (169.73,160) and (116,142.09) .. (116,120) .. controls (116,97.91) and (169.73,80) .. (236,80) -- (236,120) -- cycle ; \draw   (236,160) .. controls (236,160) and (236,160) .. (236,160) .. controls (169.73,160) and (116,142.09) .. (116,120) .. controls (116,97.91) and (169.73,80) .. (236,80) ;  
\draw    (430,80) -- (510,80) ;
\draw    (430,160) -- (510,160) ;
\draw  [line width=3]  (116,120.5) .. controls (116,120.22) and (116.22,120) .. (116.5,120) .. controls (116.78,120) and (117,120.22) .. (117,120.5) .. controls (117,120.78) and (116.78,121) .. (116.5,121) .. controls (116.22,121) and (116,120.78) .. (116,120.5) -- cycle ;
\draw  [draw opacity=0][dash pattern={on 0.84pt off 2.51pt}] (236,160) .. controls (236,160) and (236,160) .. (236,160) .. controls (224.95,160) and (216,142.09) .. (216,120) .. controls (216,97.91) and (224.95,80) .. (236,80) -- (236,120) -- cycle ; \draw  [dash pattern={on 0.84pt off 2.51pt}] (236,160) .. controls (236,160) and (236,160) .. (236,160) .. controls (224.95,160) and (216,142.09) .. (216,120) .. controls (216,97.91) and (224.95,80) .. (236,80) ;  
\draw  [draw opacity=0][dash pattern={on 0.84pt off 2.51pt}] (236,80) .. controls (236,80) and (236,80) .. (236,80) .. controls (247.05,80) and (256,97.91) .. (256,120) .. controls (256,142.09) and (247.05,160) .. (236,160) -- (236,120) -- cycle ; \draw  [dash pattern={on 0.84pt off 2.51pt}] (236,80) .. controls (236,80) and (236,80) .. (236,80) .. controls (247.05,80) and (256,97.91) .. (256,120) .. controls (256,142.09) and (247.05,160) .. (236,160) ;  
\draw  [draw opacity=0][dash pattern={on 0.84pt off 2.51pt}] (430,160) .. controls (430,160) and (430,160) .. (430,160) .. controls (418.95,160) and (410,142.09) .. (410,120) .. controls (410,97.91) and (418.95,80) .. (430,80) -- (430,120) -- cycle ; \draw  [dash pattern={on 0.84pt off 2.51pt}] (430,160) .. controls (430,160) and (430,160) .. (430,160) .. controls (418.95,160) and (410,142.09) .. (410,120) .. controls (410,97.91) and (418.95,80) .. (430,80) ;  
\draw  [draw opacity=0][dash pattern={on 0.84pt off 2.51pt}] (430,80) .. controls (430,80) and (430,80) .. (430,80) .. controls (441.05,80) and (450,97.91) .. (450,120) .. controls (450,142.09) and (441.05,160) .. (430,160) -- (430,120) -- cycle ; \draw  [dash pattern={on 0.84pt off 2.51pt}] (430,80) .. controls (430,80) and (430,80) .. (430,80) .. controls (441.05,80) and (450,97.91) .. (450,120) .. controls (450,142.09) and (441.05,160) .. (430,160) ;  
\draw  [dash pattern={on 4.5pt off 4.5pt}]  (236,80) -- (430,80) ;
\draw  [dash pattern={on 4.5pt off 4.5pt}]  (240,160) -- (430,160) ;
\draw    (430,190) -- (518,190) ;
\draw [shift={(520,190)}, rotate = 180] [color={rgb, 255:red, 0; green, 0; blue, 0 }  ][line width=0.75]    (10.93,-3.29) .. controls (6.95,-1.4) and (3.31,-0.3) .. (0,0) .. controls (3.31,0.3) and (6.95,1.4) .. (10.93,3.29)   ;
\draw    (242,190) -- (330,190) ;
\draw [shift={(240,190)}, rotate = 0] [color={rgb, 255:red, 0; green, 0; blue, 0 }  ][line width=0.75]    (10.93,-3.29) .. controls (6.95,-1.4) and (3.31,-0.3) .. (0,0) .. controls (3.31,0.3) and (6.95,1.4) .. (10.93,3.29)   ;

\draw (301,39.4) node [anchor=north west][inner sep=0.75pt]    {$S^{1} \times _{q}$};
\draw (541,110.4) node [anchor=north west][inner sep=0.75pt]  [color={rgb, 255:red, 208; green, 2; blue, 27 }  ,opacity=1 ]  {$D_{\alpha }$};
\draw (340,177.4) node [anchor=north west][inner sep=0.75pt]    {$A^{R} =\ \frac{1}{2} d\phi $};
\draw (24,105.4) node [anchor=north west][inner sep=0.75pt]    {$A^{R} =\ \frac{1}{2} w$};

\end{tikzpicture}
    
    \caption{The stretched half supconformal index in the $b \rightarrow \infty$ limit. }
    \label{fig:stretched_cigar}
\end{figure}
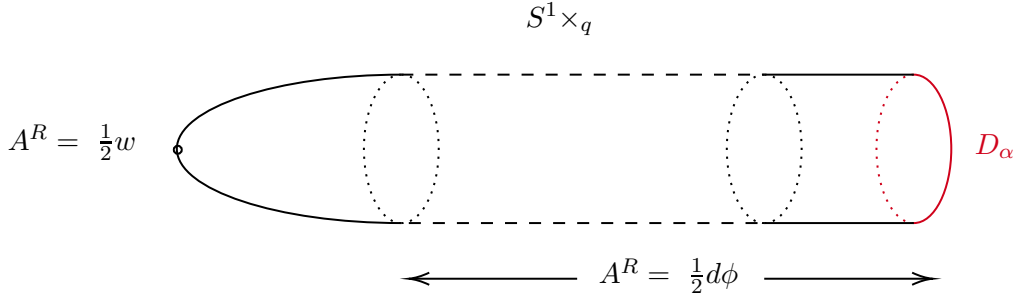

Precisely, the two indices are related via:
\begin{equation}\label{eq:tt_sc_relation}
    \mathcal{Z}(q,t,x,\xi) = \mathcal{Z}_{\text{SC}}(q, tq^{-\frac{1}{2}},x,\xi).
\end{equation}
This is not a coincidence and has been anticipated in \cite{bullimore2021boundaries, crew2020factorisation}. We now briefly explain why this is the case with an argument parallel to the 2d case \cite{Gomis:2012wy}. The rigid supersymmetry background admits a $Q$-exact squashing of $HS^2$ \cite{tanaka2015superconformal}, on which the final answer does not depend.\footnote{A more careful argument would also show that any boundary terms produced by the squashing are also $Q$-exact, however in the interest of brevity we do not show this here.} As the squashing parameter $b \rightarrow \infty$, the $HS^2$ geometry approaches that of an infinite cigar. The $R$-symmetry connection $A^R$ in the rigid supergravity background becomes equal to half the spin connection $A^R = \frac{1}{2} w$ in the region surrounding the tip, and becomes flat with a non-zero holonomy $A^R = \frac{1}{2}d\phi$ in the infinite cylindrical region, as shown in figure \ref{fig:stretched_cigar}.

Near the tip, $A^R$ implements the topological twist and at the boundary, the holonomy is equivalent to converting the boundary conditions for the fermions around $\partial HS^2$ from NS (anti-periodic) to R (periodic). The final object is then equivalent to the topologically twisted partition function on the cigar with periodic boundary conditions and therefore equal to \eqref{eq:index}. Note that the fugacity shift in \eqref{eq:tt_sc_relation} would naively produce a grading $q^{J+R_C/2}$ in \eqref{eq:index}, however in the twisted index the spins of the fields are redefined by their $R_C$ charge, so we replace this with $q^J$ where $J$ is now the action of the spin generator in the twisted theory.

\section{Localisation}\label{sec:localisation}

In this section, we compute the topologically twisted index \eqref{eq:index} with exceptional Dirichlet boundary conditions $D_{\alpha}$ using localisation. We describe two different choices of localising action that give rise to Higgs and Coulomb branch localisation schemes. We will see in section \ref{sec:quasimap} that these schemes have interesting geometric interpretations. We compute the index explicitly for supersymmetric QED; the extension to general abelian theories is straightforward. We plan to return to non-abelian theories in future work. 

We write the index as a path integral:
\begin{equation}
\mathcal{Z}_{D_{\alpha}} = \int D\Phi\exp(-S_{\text{FI}}[\Phi] - S_{\text{SQED[$N$]}}[\Phi] - t^2S_{\text{Loc.}}[\Phi]),
\end{equation}
where $\Phi$ denotes the space of fields obeying the twisted periodicities \eqref{eq:twistedperiodicity} along the time circle, and $D_{\alpha}$ on $\partial (HS^2\times S^1)$.
In the above, $S_{\text{FI}}$ denotes the Fayet-Iliopoulis term and $S_{\text{SQED}[N]}$ the sum of the vector multiplet and matter Lagrangians for SQED[$N$]. The choice of localising action $S_{\text{Loc.}} = QV$ is $Q$-exact, and thus we may send $t\to \infty$, localising the path integral to the saddle points of $QV$. The path integral will be given by the one-loop determinants of operators appearing at quadratic order in the action expanded around the saddle points, weighted by the classical action at the saddle points.  

We will study two localisation schemes corresponding to the two difference choices of localising action:
\begin{align}
\text{Coulomb:} &\quad  S_{\text{Loc.}}[\Phi]= S_{\text{SQED}[N]}[\Phi]\label{eq:coulomb_loc}.\\
\text{Higgs:} &\quad  S_{\text{Loc.}}[\Phi] = S_{\text{SQED}[N]}[\Phi] + Q\left((\bar{\lambda} +\lambda)(\bar{X}\cdot X - \bar{Y}\cdot Y - \tau)\right).\label{eq:higgs_loc}
\end{align}
Recall that the SQED[$N$] Lagrangian is $Q$-exact, and that the term $\bar{X}\cdot X - \bar{Y}\cdot Y$ is the moment map $\mu_{\mathbb{R}}$ for the $G=U(1)$ action. The Coulomb branch localising action was previously considered in \cite{benini2015topologically}, and the Higgs in \cite{bullimore2019twisted}.

In the remainder of the section we take supersymmetric QED as a running example. We no longer take vector multiplet fields to act in representations implicitly, but instead couple explicitly depending on the charges of the matter fields.

\subsection{Coulomb branch BPS locus}

The Coulomb branch localisation calculation closely follows the results of \cite{benini2015topologically}. We note first that the bosonic part of the full off-shell SQED[$N$] Lagrangian is not positive definite, so as in \textit{loc. cit.} we use the supersymmetry transformation law to determine that in the BPS locus 
\begin{align}
Q\lambda = Q\bar{\lambda} = 0 \quad \text{implies} \quad 2iF_{z\bar{z}} + D = 0.
\end{align}
Further, $\sigma$ and $A_3$ are arbitrary constants, and are set to their boundary values \eqref{eq:dirichletvector}. For nonzero $\sigma$, the potential term $\sigma^2(\bar{X}\cdot X + \bar{Y}\cdot Y)$ forces $X = Y = 0$. As a result, we must set the boundary value $c$ of the Dirichlet $\mathcal{N}=2$ chiral multiplet to zero in this localisation scheme. The Dirichlet boundary condition on $\phi_{\bar{z}}$ forces $\phi_{\bar{z}} = 0$. The vector multiplet action imposes $F_{\bar{z}3} = 0$, and, as in \cite{benini2015topologically}, one may also argue that $F_{z\bar{z}} = \text{const}$.

In summary, the BPS locus of Coulomb branch localisation consists of topologically distinct sectors:
\begin{equation}
F_{12} = -\frac{2\pi k}{A(HS^2)},\quad  A_3 + i\sigma = \text{const}, \quad X = Y = \phi_{\bar{z}} = 0,
\end{equation}
for $k \in \mathbb{Z}$. We recall that $s = \exp(i \beta(A_3  + i\sigma))$ can be arbitrary in Coulomb branch localisation, and the final result of the Coulomb branch localisation depends on it. Finally, we note that this BPS locus is compatible with the exceptional Dirichlet boundary conditions discussed in section \ref{sec:sqed_ed}.

\subsection{Higgs branch BPS locus}
We now study the BPS locus corresponding to the Higgs branch localising action \eqref{eq:higgs_loc}. The part of the gauge sector Lagrangian involving $F_{z\bar{z}}$ and $D$ is
\begin{equation}
\begin{aligned}
4&|F_{z\bar{z}}|^2 - D^2 + (2iF_{z\bar{z}} +D)(2\bar{X}\cdot X - 2\bar{Y}\cdot Y - 2\tau).
\end{aligned}
\end{equation}
Integrating out the auxiliary field $D$ and completing the square leaves
\begin{equation}
(2iF_{z\bar{z}} + \bar{X}\cdot X - \bar{Y}\cdot Y  -\tau)^2,
\end{equation}
which yields the BPS equation
\begin{equation}
2iF_{z\bar{z}} = \tau -\bar{X}\cdot X + \bar{Y}\cdot Y .
\end{equation}
Further, the matter kinetic terms impose:
\begin{equation}
\begin{aligned}
(\nabla_{\bar{z}} + iA_{\bar{z}})X &= 0, \quad 
&(D_3 + iA_3 + iA_3^t/2 + iA_3^i)X_i =0, \\
(\nabla_{\bar{z}} - iA_{\bar{z}})Y &= 0, \quad 
&(D_3 -iA_3+iA_3^t/2 - iA_3^i)Y_i =0,
\end{aligned}
\end{equation}
where we have turned on background gauge fields and masses. Combined with taking the minima of the bosonic potential\footnote{The term $(\bar{X}\cdot X - \bar{Y}\cdot Y)^2/4$ is cancelled after adding the Higgs branch localising action.} \eqref{eq:potential} and the bosonic part of the vector multiplet Lagrangian \eqref{eq:lagrangian_vector}, we arrive at the full BPS locus for the Higgs branch localisation scheme 
\begin{equation}\label{eq:higgs_bps}
\begin{aligned}
&\quad 2iF_{z\bar{z}} = \tau-\bar{X}\cdot X + \bar{Y}\cdot Y,\quad X\cdot Y = 0, \quad F_{z3} = 0, \\
&(\nabla_{\bar{z}} + iA_{\bar{z}})X = 0, \quad (D_3 + iA_3 + iA_3^t/2 + iA_3^i)X_i=0, \\
&(\nabla_{\bar{z}} - iA_{\bar{z}})Y = 0, \quad (D_3 -iA_3+iA_3^t/2 - iA_3^i)Y_i =0,\\
&\qquad\quad (\sigma + m_i + m_t/2)X_i = 0, \quad \phi_{\bar{z}}X_i = 0,\\ 
&\qquad\quad (-\sigma - m_i + m_t/2)Y_i = 0, \quad \phi_{\bar{z}}Y_i = 0, \\
&\qquad\qquad\qquad\qquad \nabla_\mu\sigma = \nabla_\mu\phi_{\bar{z}} = 0.
\end{aligned}
\end{equation}

Note that with an exceptional Dirichlet boundary condition, $X_\alpha$ is non-zero on the boundary torus. In addition, the BPS locus above implies $X_\alpha$ is covariantly holomorphic and so its zeros at constant $x^3$ must be isolated. Therefore, we must have $\sigma + m_\alpha + m_t/2 = 0$, and $\phi_{\bar{z}} = 0$ on $HS^2\times S^1$. When $m_i$ and $m_t$ are generic (as we assume), $\sigma + m_i + m_t/2$ is nowhere vanishing and thus the BPS locus requires $X_i = 0$ for all $i \neq \alpha$. Similarly, all $Y_i$ must vanish and we may therefore rewrite the Higgs branch BPS locus as follows:
\begin{equation}\label{eq:higgs_bps_reduced}
\begin{aligned}
    &\qquad 2iF_{z\bar{z}} = \tau-\bar{X}_\alpha X_\alpha, \quad F_{z3}=0,\\
    &X_i = 0 \quad \text{for } i\neq \alpha, \quad Y_i = 0 \quad \text{for all } i,\\
    &\quad(\nabla_3 + iA_3 + i A_3^\alpha + iA_3^t/2)X_\alpha = 0,\\
    &\qquad \sigma =-m_\alpha - m_t/2, \quad \phi_{\bar{z}} = 0.
\end{aligned}
\end{equation}

We may also pick a gauge for the hemisphere vortex so that the spatial part of the gauge field is $SO(2)_E$-invariant and time independent. The BPS equation $F_{z3} = 0$ implies that $A_3$ is constant and so $A_3 = -A_3^{\alpha} - A_3^t/2$ due to the boundary condition, thus $X_\alpha$ is also constant in $x^3$. In this gauge, the BPS equations may then be written as
\begin{equation}\label{eq:higgs_bps_reduced_2}
\begin{aligned}
    &\qquad\qquad\qquad\quad 2iF_{z\bar{z}} = \tau-\bar{X}_\alpha X_\alpha,\\
    &\,\,X_i = 0 \,\, \text{for } i\neq \alpha, \quad \nabla_3X_\alpha = 0, \quad  Y_i = 0 \,\,\,\text{for}\, \, i=1,\ldots, N,\\
    &\qquad A_3 = -A_3^\alpha - A_3^t/2,\quad \sigma =-m_\alpha - m_t/2, \quad \phi_{\bar{z}} = 0, \\
\end{aligned}
\end{equation}

There exists a unique solution $X_{\alpha}$ to the above equations, subject to the exceptional Dirichlet boundary conditions of section \ref{sec:sqed_ed}. We provide a sketch here and direct the reader to  \cite{manton2023neumann} for a more detailed argument.

\paragraph{Uniqueness.} Note that $X_\alpha(x^3 = 0)$ cannot have zeroes outside the origin, else the twisted periodicity $X_\alpha(X^3 = \beta) = q^{-J} X_\alpha(x^3 = 0)$ allows us to rotate $X_\alpha$ by $q$ to generate a dense set of zeros on a circle centred at the origin (we assume that $q$ is generic). The location of the zeros of $X_\alpha$ are constant in $x^3$ as implied by \eqref{eq:higgs_bps}. In the flux $k$ sector, $X_\alpha\propto \exp(ik\theta)$ on the boundary torus in a global trivialisation of the $U(1)$ bundle.\footnote{Note this is not the trivialisation provided by the exceptional Dirichlet boundary condition.} Hence $X_\alpha$ has a zero of order $k$ at the origin along each time slice. The rest of the proof is completed by showing that $h\vcentcolon=\log|X_\alpha(x^3 = \text{const})|^2$ satisfies a second order elliptic equation depending only on the zeros of $X_\alpha$ at each constant $x^3$ slice. 

\paragraph{Existence.} The existence of a solution is not guaranteed for arbitrary values of $\tau$ and the radius of the hemisphere because we have imposed both Dirichlet and Neumann boundary condition on $h$; this gives an over-determined elliptic boundary value problem. The path integral may be localised independently in each flux sector, and since it is quasi-topological in each we may prove existence under the assumption that we can change both the radius $R(HS^2)$ and $\tau$ independently in each flux sector. To prove existence in the flux $k$-sector, we first work on hemisphere with radius $1$ and take $\tau$ to be sufficiently large such that the $k$-vortex solution exists on the hemisphere. Let $h_0$ denote the boundary value of $h$ on this hemisphere, the vortex equations \eqref{eq:higgs_bps_reduced} are then invariant under the transformation
\begin{equation}
X_\alpha\to \frac{c}{h_0}X_\alpha, \quad \tau \to \frac{c^2}{h_0^2}\tau,\quad  R(HS^2)\to \frac{h_0}{c}R(HS^2),
\end{equation}
and the new solution satisfies $|X_\alpha| = c$ on the boundary and therefore satisfies the exceptional Dirichlet boundary condition of section \ref{sec:sqed_ed}.

\subsection{Localisation}\label{subec:localisation}

In this section we present the results of the localisation procedure. The arguments in the following hold for both Coulomb and Higgs branch localisation since the only conditions required are that $A_3$ and $\sigma$ are constant and the spatial part of the gauge field is $SO(2)_E$ invariant. These conditions are satisfied by both vortex and monopole backgrounds. As described previously, the Coulomb branch localisation will depend on an additional fugacity $s=e^{i\beta(A_3+i\sigma)|_{\partial}}$ whereas in the Higgs branch localisation, $s$ is specialised according the values of $A_3$ and $\sigma$ in the vacuum. For supersymmetric QED, this is given explicitly by $s x_{\alpha}t^{\frac{1}{2}}=1$ (which follows from either \eqref{eq:substitution_fugacity} or the weight description in section \ref{subsec:higgsbranch}). 

We begin by deriving one-loop determinants for $\mathcal{N}=2$ multiplets before turning to $\mathcal{N}=4$ theories and hence products thereof. We compute the one-loop determinants in the following sections by eigenvalue-matching, an alternative method of computing one-loop determinants by rewriting them as equivariant indices is reviewed in appendix \ref{ap:equivariant_index}.

\subsubsection{\texorpdfstring{$\mathcal{N}=2$}{} Dirichlet chiral}
In this section we compute the one-loop determinant of a charge $1$ chiral multiplet in a flux $k$ background with Dirichlet boundary conditions \eqref{eq:chiral_boundary_conditions}. We work in the presence of a background connection $A_3^{(B)}$ and real mass $m$.  

The quadratic action for chiral multiplet fluctuations relevant for this computation is
\begin{equation}
  \begin{aligned}
&D_3\delta\bar{X}D_3 \delta X + 4D_z \delta\bar{X}D_{\bar{z}} \delta X + (\sigma + m)^2|\delta X|^2   + 2\bar{\psi}D_z\psi_{\bar{z}}  \\
&\quad-2\bar{\psi}_z\nabla_{\bar{z} }\psi -\bar{\psi}(D_3 + \sigma + m)\psi - \bar{\psi}_z(D_3 - \sigma - m)\psi_{\bar{z}},
\end{aligned}  
\end{equation}
where we denote the gauge covariant derivative  $D = \nabla + iA + iA^{(B)}$. The other terms in the chiral multiplet action (such as the Yukawa coupling) do not contribute to the one-loop determinant. The Dirichlet boundary conditions imply that the one-loop fluctuations satisfy
\begin{equation}
\delta X = \delta\bar{X} = \psi = \bar{\psi} = 0,
\end{equation}
at the boundary and must in addition obey the twisted periodicities \eqref{eq:twistedperiodicity}.
The one-loop determinant is then given by
\begin{equation}
\frac{\det(\slashed{D} + \sigma + m)}{\det(-4D_zD_{\bar{z}} - D_3D_3+(\sigma + m)^2)}.
\end{equation}
We match the eigenvalues of the 3d twisted Dirac operator with eigenvalues of the 3d Laplacian. An eigenfunction of the 3d Laplacian has the form
\begin{equation}
\delta X \propto X_{\text{2d}}\exp((2\pi in + j\epsilon) x^3/\beta),\quad  n\in\mathbb{Z}, \,j\in\mathbb{Z},
\end{equation}
where $X_{2d}$ is an eigenfunction of the 2d Laplacian with Dirichlet boundary condition
\begin{equation}
-4D_zD_{\bar{z}}X_{\text{2d}} = \lambda_{\text{2d}} X_{\text{2d}},\qquad X_{\text{2d}}|_{\partial HS^2} = 0,
\end{equation}
and $j$ is the angular momentum of $X_{\text{2d}}$: $X_{\text{2d}}(qw) = q^{j+k} X_{\text{2d}}(w)$ with $k$ the magnetic flux. The eigenvalue of the 3d Laplacian together with the $(\sigma+m)^2$ term is then
\begin{equation}
\lambda_{\text{2d}} - ((2\pi in + j\epsilon)/\beta + iA_3 + iA_3^{(B)})^2 + (\sigma + m)^2.
\end{equation}

We will show that this eigenvalue is cancelled by the determinant of the twisted Dirac operator on the following two dimensional subspace of fermions:
\begin{equation}
\begin{aligned}
\begin{pmatrix}
    \psi \\ \psi_{\bar{z}}
\end{pmatrix}
\in
\left\langle
\begin{pmatrix}
X_{\text{2d}}\exp((2\pi in + j\epsilon)x^3/\beta)\\
0
\end{pmatrix},
\begin{pmatrix}
0\\
2D_{\bar{z}}X_{\text{2d}}\exp((2\pi in + j\epsilon)x^3/\beta)
\end{pmatrix}
\right\rangle
\end{aligned}
\end{equation}
The matrix elements of the Dirac operator:
\begin{align}
\begin{pmatrix}
D_3 + \sigma + m & -2D_z\\
2D_{\bar{z}} &D_3 - \sigma - m
\end{pmatrix},
\end{align}
acting on space of fermions $(\psi,\psi_{\bar{z}})$ in this basis are:
\begin{equation}
    \begin{pmatrix}
(2\pi in + j\epsilon)/\beta+ iA_3+iA_3^{(B)} + \sigma+m& \lambda_{\text{2d}}\\
1 &  (2\pi in + j\epsilon)/\beta + iA_3 +i A_3^{(B)}-\sigma-m
\end{pmatrix}.
\end{equation}
Its determinant then cancels the 3d Laplacian eigenvalue. We now consider the unmatched modes.

Since $X$ obeys a Dirichlet boundary condition there are no bosonic unmatched modes with $\lambda_{\text{2d}} = 0$. However, $\psi_{\bar{z}}$ has zero modes because the map $\nabla_{\bar{z}}$ from $\psi$ to $\psi_{\bar{z}}$ is not surjective. Its cokernel is the space of $\psi_{\bar{z}}$ which satisfies
\begin{equation}
\int_{HS^2} D_z\psi \psi_{\bar{z}} = 0, \quad \text{for all } \psi.
\end{equation}
Integrating by parts and noting that the boundary term vanishes due to the boundary condition on $\psi = 0$, we deduce that $\psi_{\bar{z}}$ is anti-holomorphic. The unmatched modes of $\psi_{\bar{z}}$ are then
\begin{equation}
\psi_{\bar{z}}d\bar{z} = \psi_{\bar{z}}^0 \bar{w}^l\exp((2\pi in - (l+k + 1)\epsilon) x^3/\beta),\quad l \in \mathbb{Z}_{\geq 0}, n\in\mathbb{Z},
\end{equation}
where $\psi_{\bar{z}}^0$ is a nowhere vanishing covariantly holomorphic $(0,1)$ form\footnote{To argue the existence of this state, we first discuss the case when $SO(2)_E$ acts trivially on the fibre above zero. We may then choose an arbitrary holomorphic section $f_0$ which is non-zero at the origin. We define $\tilde{F} = \int_{SO(2)_E} g^*f_0dg $ which is the pull back of $f_0$ by an element $g$ of $SO(2)_E$ averaged over $SO(2)_E$. We observe $\tilde{F}(0) = F_0(0)\neq 0$ and is $SO(2)_E$-equivariant. This implies that $\tilde{F}$ is nowhere vanishing since if it vanishes at some point then it must vanish on a circle and $\tilde{F}$ is non-zero. The general case with arbitrary $k$ can be reduced to this case by taking a tensor product with the trivial holomorphic bundle with weight $-k$ under $SO(2)_E$.} valued in the charge $1$ (under $G$) line bundle with angular momentum $-k-1$ and the coordinate $w$ is the coordinate on $HS^2$ defined at the beginning of section \ref{sec:hem_geom}.
These modes contribute:
\begin{equation}\label{dirichlet_one_loop}
\begin{aligned}
\text{det}&(\nabla_3 + iA_3 + iA_3^{(B)} - \sigma-m)\\
    &= \prod_{n\in\mathbb{Z}, l \in \mathbb{Z}_{\geq 0}} 2\pi in - (l + k+1)\epsilon +i\beta ( A_3^B+  \beta A_3) - \beta(\sigma + m) \\
    &= \,\,\left(\prod_{j\geq 0}s^{1/2}q^{-(1+k)/2}x_f^{1/2}q^{-j/2}\right) (s^{-1}q^{1 + k}x_f^{-1};q)_\infty\\
    &=\exp(-\log(s^{-1/2}q^{(1+k)/2}x_f^{-1/2})/2+\log(s^{-1/2}q^{(1+k)/2}x_f^{-1/2})^2/ \log q) \\
    &\quad\,(s^{-1}q^{1 + k}x_f^{-1};q)_\infty
\end{aligned}
\end{equation}
In the above we have introduced complexified fugacities $x_f = \exp(i\beta A_3^B - \beta m)$ and $s = \exp(i\beta A_3 - \beta \sigma)$ and made use of the zeta function regularisation:
\begin{equation}
\prod_{j=0}^\infty (yq^j) = \exp(\frac{1}{2}\log(y) - \frac{\log(y)^2}{2\log(q)})
\end{equation}
as in \cite{yoshida2014localization,bullimore2021boundaries, tanaka2015superconformal}. Note that we have omitted an overall constant in the exponential since it will later cancel in the 3d $\mathcal{N}=4$ one-loop determinants. The exponential factor from the regularisation is closely related to boundary 't Hooft anomalies arising from the chiral with this boundary condition:, as elucidated in \cite{bullimore2016boundaries}. This is true for all other one-loop determinants in this work: the prefactor is determined precisely by the boundary anomaly polynomial. Later in this work we give another interpretation of these prefactors in terms of Higgs branch geometry.

\subsubsection{\texorpdfstring{$\mathcal{N}=2$}{} Neumann chiral}\label{sec:neumann_chiral}
The computation of the one-loop determinant of a charge $1$ Neumann chiral mostly follows that of a charge one Dirichlet chiral. A difference is that to match eigenvalues one must choose an orthonormal basis for the 2d bosonic Laplacian; whilst the existence of an orthonormal basis for the 2d Dirichlet Laplacian is standard, the Neumann case is perhaps less well known and so we provide an argument for the existence of such an orthonormal basis in appendix \ref{ap:eigenvalue_neumann}. 

The matching of eigenvalues is the same as for the Dirichlet case, so we only study the unmatched modes in the following. The Neumann boundary condition in a vanishing gaugino background is
\begin{equation}
D_{\bar{z}}X = D_z\bar{X}= D_{\bar{z}}\psi = D_z\bar{\psi} = D_{\bar{z}}\bar{\psi}_z = D_z\psi_{\bar{z}} = 0.
\end{equation}
Only $\psi$ and $X$ have unmatched modes since the map $D_{\bar{z}}$ from $\psi$ to $\psi_{\bar{z}}$ is now surjective and so $\psi_{\bar{z}}$ does not have any unmatched modes. $\psi$ is manifestly covariantly holomorphic, as is  $X$ since 
$D_zD_{\bar{z}}X = 0$ and $D_{\bar{z}}X = 0|_\partial$ imply $D_{\bar{z}}X = 0$ everywhere. In summary, the unmatched modes are:
\begin{align}
X &= X^0 w^l \exp((2\pi in + (l-k)\epsilon) x^3/\beta), \quad \text{for } l \in \mathbb{Z}^{\ge 0},n\in\mathbb{Z},\\
\psi &= X^0 w^l \exp((2\pi in + (l-k)\epsilon) x^3/\beta), \quad \text{for } l \in \mathbb{Z}^{\ge 0},n\in\mathbb{Z},
\end{align}
where $X^0$ is a global non-vanishing holomorphic section of the charge one line bundle with angular momentum $-k$. Their contribution to the one-loop determinant is given by
\begin{equation}\label{neumann_one-loop}
\begin{aligned}
&\frac{\det (D_3 + \sigma+m)}{\det ((D_3 - \sigma-m)(D_3 + \sigma+m))}=\frac{1}{\det (D_3 - \sigma - m)}\\
&\quad= \prod_{ n\in\mathbb{Z},\,l \in \mathbb{Z}_{\geq 0}}(2\pi in + (l-k)\epsilon+i\beta A_3+i\beta A_3^{(B)} -\beta \sigma - \beta m)^{-1}\\
&\quad= \Bigg(\prod_{j\geq 0} s^{1/2}x_f^{1/2}q^{-k/2}q^{j/2}\Bigg)(sx_fq^{-k} ;q)^{-1}_{\infty}\\
&\quad= \exp(\log(s^{1/2}x_f^{1/2}q^{-k/2})/2 - \log^2(s^{1/2}x_f^{1/2}q^{-k/2})/\log q)(sx_fq^{-k};q)_{\infty}^{-1}.
\end{aligned}
\end{equation}

\subsubsection{\texorpdfstring{$\mathcal{N}=2$}{} Dirichlet vector}
We now compute the one loop determinant of an abelian $\mathcal{N}=2$ Dirichlet vector multiplet. The quadratic fluctuation of the vector multiplet Lagrangian is
\begin{equation}
\frac{1}{2}\delta F_{\mu\nu}\delta F^{\mu\nu} + \nabla_\mu\delta\sigma\nabla^\mu\delta\sigma +\bar{\lambda}\nabla_3\lambda +\bar{\lambda}_z\nabla_3\lambda_{\bar{z}}+2\bar{\lambda}_z\nabla_{\bar{z}}\lambda-2\bar{\lambda}\nabla_z\lambda_{\bar{z}}.
\end{equation}
In order to compute the one-loop determinant, the fluctuation $\delta A_\mu$ must be gauge fixed. The gauge fixed Lagrangian with BRST ghosts is:
\begin{align}
&\frac{1}{2}\delta F_{\mu\nu}\delta F^{\mu\nu} + \nabla^\mu\delta\sigma\nabla_\mu\delta\sigma\nonumber+ \nabla_\mu\delta A^\mu \nabla_\nu \delta A^\nu - \bar{C}\nabla_\mu\nabla^\mu C\nonumber\\
&+\bar{\lambda}\nabla_3\lambda +\bar{\lambda}_z\nabla_3\lambda_{\bar{z}}+2\bar{\lambda}_z\nabla_{\bar{z}}\lambda-2\bar{\lambda}\nabla_z\lambda_{\bar{z}}\nonumber\\
=&-\delta A^j\nabla^\mu\nabla_\mu \delta A_j - (\delta A_3 - i\delta \sigma)\nabla_\mu\nabla^\mu (\delta A_3 + i\delta \sigma) - \bar{C}\nabla_\mu\nabla^\mu C\nonumber\\
&+\bar{\lambda}\nabla_3\lambda +\bar{\lambda}_z\nabla_3\lambda_{\bar{z}}+2\bar{\lambda}_z\nabla_{\bar{z}}\lambda-2\bar{\lambda}\nabla_z\lambda_{\bar{z}},
\end{align}
where $i,j$ are spatial indices taking values $1,2$. 

We now impose the boundary conditions $C = \bar{C} = 0$.\footnote{$C$ generates gauge transformations so that in order to preserve the boundary condition $A_3 = A_t= \text{const.}$, $C$ must be constant on the boundary. We eliminate a zero mode of $C$ by setting it to zero on the boundary.} We note that the contribution from the ghost cancels the contribution from $A_3 + i\sigma$ and so it remains to compute the contribution from $A_i$ and the fermions. The fermion kinetic term may be written as
\begin{equation}
\frac{1}{2}
\begin{pmatrix}
\lambda &\lambda_{\bar{z}}&\bar{\lambda}&\bar{\lambda}_z
\end{pmatrix}
\begin{pmatrix}
0&0&\nabla_3&2\nabla_{\bar{z}}\\
0&0&-2\nabla_z&\nabla_3\\
\nabla_3&-2\nabla_z&0&0\\
2\nabla_{\bar{z}}&\nabla_3&0&0
\end{pmatrix}
\begin{pmatrix}
\lambda\\
\lambda_{\bar{z}}\\
\bar{\lambda}\\
\bar{\lambda}_z
\end{pmatrix},
\end{equation}
and the bosonic kinetic term for $A_i$ is $-\nabla_\mu\nabla^\mu + R$, where $R$ denotes the Ricci tensor. The one-loop determinant of the vector multiplet may be expressed as:\footnote{The Pffafian of the fermion differential operator is the square root of its determinant.}
\begin{equation}
\text{det}^{\frac{1}{2}}\begin{pmatrix}
0&0&\nabla_3&2\nabla_{\bar{z}}\\
0&0&-2\nabla_z&\nabla_3\\
\nabla_3&-2\nabla_z&0&0\\
2\nabla_{\bar{z}}&\nabla_3&0&0
\end{pmatrix} \left( \text{det}^{\frac{1}{2}}(-\nabla_\mu\nabla^\mu + R) \right)^{-1}.
\end{equation}
Starting with a 2d eigenmode $A_i^{\text{2d}}$ of the vector Laplacian
\begin{equation}\label{vector_laplacian_eigenfunction}
-\nabla^i\nabla_i A_j^{\text{2d}} + R_i^jA_i^{\text{2d}} = \kappa^2 A_j^{\text{2d}},
\end{equation}
or equivalently
\begin{equation}
-4\nabla_z\nabla_{\bar{z}}A_z^{\text{2d}} = \kappa^2A_z^{\text{2d}},
\end{equation}
we may construct a 2d fermionic eigenmode
\begin{equation}
\begin{pmatrix}
\lambda\\
\lambda_{\bar{z}}\\
\bar{\lambda}\\
\bar{\lambda}_z
\end{pmatrix}
\propto
\begin{pmatrix}
-2\nabla_zA_{\bar{z}}^{\text{2d}}\\
\kappa A_{\bar{z}}^{\text{2d}}\\
2\nabla_{\bar{z}}A_z^{\text{2d}}\\
\kappa A_z^{\text{2d}}
\end{pmatrix},
\end{equation}
which satisfies
\begin{equation}
\begin{pmatrix}
0&0&0&2\nabla_{\bar{z}}\\
0&0&-2\nabla_z&0\\
0&-2\nabla_z&0&0\\
2\nabla_{\bar{z}}&0&0&0
\end{pmatrix}
\begin{pmatrix}
\lambda\\
\lambda_{\bar{z}}\\
\bar{\lambda}\\
\bar{\lambda}_z
\end{pmatrix}
=\kappa
\begin{pmatrix}
0&0&1&0\\
0&0&0&1\\
1&0&0&0\\
0&1&0&0
\end{pmatrix}
\begin{pmatrix}
\lambda\\
\lambda_{\bar{z}}\\
\bar{\lambda}\\
\bar{\lambda}_z
\end{pmatrix}.
\end{equation}
Due to the symmetry  $\kappa\to  -\kappa$ a bosonic mode is matched with two fermionic modes. Similarly, starting with a 2d eigenmode $A_i^{\text{2d}}$ we may construct a 3d eigenmode
\begin{equation}
A_i\propto A_i^{\text{2d}}\exp((2\pi in + j\epsilon)x^3/\beta), \quad j\in\mathbb{Z}, \,n\in\mathbb{Z},
\end{equation}
and similarly for the fermions. One can then check that the bosonic 3d Laplacian cancels the fermionic Dirac operator.

Now let us turn to the unmatched modes. The 2d bosonic vector Laplacian $-\nabla_i\nabla^i + R$ has a non-trivial kernel because the value of $A_i$ can be prescribed arbitrarily on the boundary circle provided it is normal to the boundary. Thus we have a sequence of unmatched bosonic modes $A^{p,\text{2d}}$ whose boundary values are given by $\hat{n}\exp(ip\theta)$ where $\hat{n}$ is the outward unit normal vector, $\theta$ is the angular coordinate on the boundary, and $p \in \mathbb{Z}$ is the angular momentum. The unmatched 2d fermionic modes are given by:
\begin{equation}
\begin{pmatrix}
\lambda\\
\lambda_{\bar{z}}\\
\bar{\lambda}\\
\bar{\lambda}_z
\end{pmatrix}
\propto
\begin{pmatrix}
-2\nabla_zA^{p,\text{2d}}_{\bar{z}}\\
0\\
2\nabla_{\bar{z}}A^{p,\text{2d}}_z\\
0
\end{pmatrix}.
\end{equation}
Combining, we find the one-loop determinant for the vector multiplet is
\begin{equation}\label{vector_one_loop}
\begin{aligned}
\frac{1}{(\text{det} D_3)^{\frac{1}{2}}} &= \prod_{n\in\mathbb{Z}\,, p\in\mathbb{Z}}(2\pi in + p\epsilon)^{-\frac{1}{2}} \\
&= \exp(\log(q^{1/2})/2-\log(q^{1/2})^2/\log q) (q;q)^{-1}_{\infty}.
\end{aligned}
\end{equation}

\paragraph{$\mathcal{N}=2$ adjoint chiral multiplet.}

The computation of the one-loop determinant of the adjoint chiral is similar to the one-loop determinant of the vector multiplet. For application to 3d $\mathcal{N}=4$ theories, we assume that the adjoint chiral has charge $1$ under a symmetry $T_t$ with background connection $A_3^{(t)}$ and real mass $m_t$. We introduce a corresponding complexified fugacity $t = \exp(i\beta A_3^{(t)} - \beta m_t)$. The matching of eigenmodes is similar to that of a free Dirichlet chiral. We note first that $\eta_{\bar{z}}, \phi_{\bar{z}}$ do not have unmatched modes. The unmatched modes of $\eta$ are
\begin{equation}
\eta \propto \eta^0w^p\exp(2\pi in + p\epsilon)\,, \quad p \in \mathbb{Z}_{\ge 0},\, n\in\mathbb{Z},
\end{equation}
where $\eta^0$ is a global $SO(2)_E$-invariant holomorphic section of the adjoint bundle.
The one-loop determinant receives the following contribution from these zero modes
\begin{equation}\label{adjoint_one_loop}
\prod_{n\in\mathbb{Z}, p\geq 0}(2\pi in + p\epsilon + i\beta A_3^{(t)} - i\beta m_t) = \frac{(t;q)_\infty}{\exp(\frac{1}{2}\log(t^{1/2}) - \log^2(t^{1/2})/\log q)}.
\end{equation}

\subsection{\texorpdfstring{$\mathcal{N}=4$}{} free hypermultiplet}

In this section we study the index of a free hypermultiplet $(X,Y)$. This theory has a $G_H = U(1)_x$ flavour symmetry. We impose Neumann boundary conditions for $X$ and a Dirichlet boundary condition for $Y$, which is an exceptional Dirichlet boundary condition (see section \ref{subsec:ed}) for the chamber $m>0$, where $m$ is a real mass for $G_H$. The hypermultiplets $X$ and $Y$ have charges $(\pm1, +1)$ under $G_H$ and $T_t$ respectively. 
The partition function is then simply the product of the two one-loop determinants
\begin{equation}
\mathcal{Z}_{\text{hyper}} 
=\exp(-\frac{\log(t/q)\log(x)/2}{\log q})\frac{(xt^{-1/2}q;q)_\infty}{(xt^{1/2};q)_\infty},
\end{equation}
where we have simplified the prefactors from the zeta function regularisation. Note that the final exponential prefactor encodes the mixed $U(1)_x-U(1)_C$ boundary 't Hooft anomaly \cite{dimofte2018dual} present for this choice of boundary condition for $(X,Y)$.

\subsection{\texorpdfstring{$\mathcal{N}=4$}{} supersymmetric QED}

We now return to our running example of $\mathcal{N}=4$ supersymmetric QED with $N$ hypermultiplets. We study the index of the exceptional Dirichlet boundary condition $D_{\alpha}$ introduced in section \ref{sec:sqed_ed}.

The result of the Coulomb branch localisation computation is
\begin{equation}\label{eq:CBresult}
\begin{aligned}
\mathcal{Z}_{D_\alpha}(s,\xi,x_i,t,q) 
= e^{\phi_{\alpha}(s)} \sum_{k\in\mathbb{Z}} \xi^k\frac{(t;q)_\infty}{(q;q)_\infty}\prod_{i\leq \alpha}\frac{(s^{-1}x_i^{-1}q^{1+k}t^{-1/2};q)_\infty}{(s^{-1}x_i^{-1}t^{1/2}q^k;q)_\infty}\prod_{i> \alpha}\frac{(sx_it^{-1/2}q^{1-k};q)_\infty}{(sx_it^{1/2}q^{-k};q)_\infty}.
\end{aligned}
\end{equation}
The index is a sum over flux sectors $k$, which may be interpreted as either the monopole flux in the Coulomb branch scheme or, upon specialising the gauge fugacities, the vortex flux in the Higgs branch scheme. The summand arises by combining the one-loop determinants \eqref{dirichlet_one_loop}, \eqref{neumann_one-loop}, \eqref{vector_one_loop} and \eqref{adjoint_one_loop} appropriate for the boundary condition $D_{\alpha}$.

In the above, there is also a classical contribution from the bulk and boundary FI terms. The bulk term \eqref{eq:3d_FI} is proportional to the vortex number $k$, and contributes $\xi^k$ to the index. The exponential prefactor arises from the boundary FI term \eqref{eq:2d_FI}
\begin{equation}
    \phi_0(s) = \frac{\log \xi \log(s^{-1})}{i\beta} = \frac{\log \xi\log(s^{-1})}{\log q}
\end{equation}
together with the contribution from the zeta function regularisation of the one-loop determinants so that in total we obtain
\begin{equation}
\begin{aligned}
    \phi_{\alpha}(s) =&\, \phi_{0}(s) + \frac{1}{4}\log(q/t) + \frac{\log(t^{1/2}q^{-1/2})\log(t^{1/2}q^{1/2})}{\log q}\\
    &+ \sum_{i\leq \alpha} \left(-\frac{1}{4}\log(q/t) + \frac{\log(q^{1/2}t^{-1/2})\log(s^{-1}x_i^{-1}q^{k+1/2})}{\log q}\right)\\
    &+ \sum_{i>\alpha} \left( -\frac{1}{4}\log(q/t)+\frac{\log(q^{1/2}t^{-1/2})\log(sx_iq^{-k+1/2})}{\log q}\right).
\end{aligned}
\end{equation}
In the above two equations we have set $q = \exp(i\beta)$, and will do in the following, in order to match the superconformal index \cite{bullimore2021boundaries}.

Upon substituting $s = s_{\alpha}= x_\alpha^{-1}t^{-1/2}$ to set the vev of $A_3 + i\sigma$ on the boundary to the value required for Higgs branch localisation, the index may be further organised as\footnote{In this expression we re-organise the $q$-Pochhammer products using standard identities and note that only the $k \ge 0$ part of the summation contributes to the index.}
\begin{equation}\label{eq:localisationsummary}
    \mathcal{Z}_{D_\alpha}(x_i,\xi,q,t) = \mathcal{Z}_{\alpha}^{\text{Classical}} \mathcal{Z}_{\alpha}^{\text{1-loop}} \mathcal{Z}_{\alpha}^{\text{Vortex}} \,.
\end{equation}
The classical piece is given by
\begin{equation}
    \mathcal{Z}_{\alpha}^{\text{Classical}} = e^{\phi_{\alpha}} \,,
\end{equation}
with
\begin{equation}
\begin{aligned}
\phi_{\alpha} = & \frac{\log \xi \log(x_{\alpha}t^{1/2})}{\log q} + (2\alpha - N - 1)\frac{\log(q^{1/2}t^{-1/2})\log(t^{1/2})}{\log q}\\
&+\sum_{i<\alpha} \frac{\log(q^{1/2}t^{-1/2})\log(x_\alpha/x_i)}{\log q} + \sum_{i>\alpha}\frac{\log(q^{1/2}t^{-1/2})\log(x_i/x_\alpha)}{\log q} \,.
\end{aligned}
\end{equation}
The one loop contribution is 
\begin{equation}\label{eq:sqed-1loop}
    \mathcal{Z}^{\text{1-loop}}_\alpha = \prod_{i< \alpha}\frac{(qx_\alpha x_i^{-1};q)_\infty}{(tx_\alpha x_i^{-1};q)_\infty}\prod_{i>\alpha}\frac{(qt^{-1}x_i x_\alpha^{-1};q)_\infty}{(x_i x_\alpha^{-1};q)_\infty} \,,
\end{equation}
and the vortex contribution is 
\begin{equation}\label{eq:vortex_pfn}
    \mathcal{Z}^{\text{vortex}}_\alpha = \sum_{k\geq 0}\xi^k(qt^{-1})^{Nk/2}\prod_{i=1}^N \frac{(tx_\alpha x_i^{-1};q)_k}{(qx_\alpha x_i^{-1};q)_k} \,.
\end{equation}
We note that, in accordance with the argument of section \ref{subsubsubsec:halfindexrelation}, we indeed recover the half superconformal index computed in §3.3 of \cite{bullimore2021boundaries} after the appropriate shift of the $t$ fugacity. In the following section we will realise each term in this index in terms of the geometry of the Higgs branch $\mathcal{M}_H$.

\section{Vortices and quasimap geometry}\label{sec:quasimap}

We now elucidate the equivalence between the twisted index on the hemisphere with exceptional Dirichlet boundary conditions $\mathcal{Z}_{D_{\alpha}}$ and the vertex function $\mathsf{V}_{\alpha}$ of Okounkov \textit{et. al.} (we refer the reader to \cite{okounkov2015lectures} for an excellent review of enumerative geometry and quasimaps), which is an equivariant count of based quasimaps to the Higgs branch. We show that they are related to the hemisphere path integral up to perturbative contributions, for which we give a precise geometrical interpretation. We demonstrate the relation explicitly by comparing the field theory localisation of the hemisphere twisted index with the localisation computation of the vertex function, but before doing so we motivate the connection with a physical argument.
 
\begin{figure}[ht!]
    \centering
\tikzset{every picture/.style={line width=0.75pt}} 

\begin{tikzpicture}[x=0.75pt,y=0.75pt,yscale=-0.75,xscale=0.75]

\draw  [draw opacity=0][dash pattern={on 0.84pt off 2.51pt}] (580,361) .. controls (580,361) and (580,361) .. (580,361) .. controls (568.95,361) and (560,343.09) .. (560,321) .. controls (560,298.91) and (568.95,281) .. (580,281) -- (580,321) -- cycle ; \draw  [color={rgb, 255:red, 208; green, 2; blue, 27 }  ,draw opacity=1 ][dash pattern={on 0.84pt off 2.51pt}] (580,361) .. controls (580,361) and (580,361) .. (580,361) .. controls (568.95,361) and (560,343.09) .. (560,321) .. controls (560,298.91) and (568.95,281) .. (580,281) ;  
\draw  [draw opacity=0] (580,281) .. controls (580,281) and (580,281) .. (580,281) .. controls (580,281) and (580,281) .. (580,281) .. controls (591.05,281) and (600,298.91) .. (600,321) .. controls (600,343.09) and (591.05,361) .. (580,361) -- (580,321) -- cycle ; \draw  [color={rgb, 255:red, 208; green, 2; blue, 27 }  ,draw opacity=1 ] (580,281) .. controls (580,281) and (580,281) .. (580,281) .. controls (580,281) and (580,281) .. (580,281) .. controls (591.05,281) and (600,298.91) .. (600,321) .. controls (600,343.09) and (591.05,361) .. (580,361) ;  
\draw  [draw opacity=0] (140,160) .. controls (140,160) and (140,160) .. (140,160) .. controls (73.73,160) and (20,142.09) .. (20,120) .. controls (20,97.91) and (73.73,80) .. (140,80) -- (140,120) -- cycle ; \draw   (140,160) .. controls (140,160) and (140,160) .. (140,160) .. controls (73.73,160) and (20,142.09) .. (20,120) .. controls (20,97.91) and (73.73,80) .. (140,80) ;  
\draw    (140,80) -- (280,80) ;
\draw    (140,160) -- (280,160) ;
\draw  [dash pattern={on 0.84pt off 2.51pt}]  (230,120) -- (307,120) ;
\draw [shift={(310,120)}, rotate = 180] [fill={rgb, 255:red, 0; green, 0; blue, 0 }  ][line width=0.08]  [draw opacity=0] (8.93,-4.29) -- (0,0) -- (8.93,4.29) -- cycle    ;
\draw    (380,118.5) -- (413,118.5)(380,121.5) -- (413,121.5) ;
\draw [shift={(420,120)}, rotate = 180] [color={rgb, 255:red, 0; green, 0; blue, 0 }  ][line width=0.75]    (10.93,-4.9) .. controls (6.95,-2.3) and (3.31,-0.67) .. (0,0) .. controls (3.31,0.67) and (6.95,2.3) .. (10.93,4.9)   ;
\draw   (470,120) .. controls (470,81.34) and (501.34,50) .. (540,50) .. controls (578.66,50) and (610,81.34) .. (610,120) .. controls (610,158.66) and (578.66,190) .. (540,190) .. controls (501.34,190) and (470,158.66) .. (470,120) -- cycle ;
\draw  [line width=3]  (540,60.5) .. controls (540,60.22) and (540.22,60) .. (540.5,60) .. controls (540.78,60) and (541,60.22) .. (541,60.5) .. controls (541,60.78) and (540.78,61) .. (540.5,61) .. controls (540.22,61) and (540,60.78) .. (540,60.5) -- cycle ;
\draw  [color={rgb, 255:red, 208; green, 2; blue, 27 }  ,draw opacity=1 ][line width=3]  (540,180.5) .. controls (540,180.22) and (540.22,180) .. (540.5,180) .. controls (540.78,180) and (541,180.22) .. (541,180.5) .. controls (541,180.78) and (540.78,181) .. (540.5,181) .. controls (540.22,181) and (540,180.78) .. (540,180.5) -- cycle ;
\draw  [line width=3]  (20,120.5) .. controls (20,120.22) and (20.22,120) .. (20.5,120) .. controls (20.78,120) and (21,120.22) .. (21,120.5) .. controls (21,120.78) and (20.78,121) .. (20.5,121) .. controls (20.22,121) and (20,120.78) .. (20,120.5) -- cycle ;
\draw  [draw opacity=0] (140,360) .. controls (140,360) and (140,360) .. (140,360) .. controls (73.73,360) and (20,342.09) .. (20,320) .. controls (20,297.91) and (73.73,280) .. (140,280) -- (140,320) -- cycle ; \draw   (140,360) .. controls (140,360) and (140,360) .. (140,360) .. controls (73.73,360) and (20,342.09) .. (20,320) .. controls (20,297.91) and (73.73,280) .. (140,280) ;  
\draw    (170,280) -- (290,280) ;
\draw    (170,360) -- (290,360) ;
\draw  [dash pattern={on 0.84pt off 2.51pt}]  (230,320) -- (307,320) ;
\draw [shift={(310,320)}, rotate = 180] [fill={rgb, 255:red, 0; green, 0; blue, 0 }  ][line width=0.08]  [draw opacity=0] (8.93,-4.29) -- (0,0) -- (8.93,4.29) -- cycle    ;
\draw  [line width=3]  (20,320.5) .. controls (20,320.22) and (20.22,320) .. (20.5,320) .. controls (20.78,320) and (21,320.22) .. (21,320.5) .. controls (21,320.78) and (20.78,321) .. (20.5,321) .. controls (20.22,321) and (20,320.78) .. (20,320.5) -- cycle ;
\draw  [draw opacity=0] (170,280) .. controls (170,280) and (170,280) .. (170,280) .. controls (181.05,280) and (190,297.91) .. (190,320) .. controls (190,342.09) and (181.05,360) .. (170,360) -- (170,320) -- cycle ; \draw   (170,280) .. controls (170,280) and (170,280) .. (170,280) .. controls (181.05,280) and (190,297.91) .. (190,320) .. controls (190,342.09) and (181.05,360) .. (170,360) ;  
\draw  [draw opacity=0] (140,280) .. controls (140,280) and (140,280) .. (140,280) .. controls (151.05,280) and (160,297.91) .. (160,320) .. controls (160,342.09) and (151.05,360) .. (140,360) -- (140,320) -- cycle ; \draw   (140,280) .. controls (140,280) and (140,280) .. (140,280) .. controls (151.05,280) and (160,297.91) .. (160,320) .. controls (160,342.09) and (151.05,360) .. (140,360) ;  
\draw  [draw opacity=0][dash pattern={on 0.84pt off 2.51pt}] (140,360) .. controls (140,360) and (140,360) .. (140,360) .. controls (128.95,360) and (120,342.09) .. (120,320) .. controls (120,297.91) and (128.95,280) .. (140,280) -- (140,320) -- cycle ; \draw  [dash pattern={on 0.84pt off 2.51pt}] (140,360) .. controls (140,360) and (140,360) .. (140,360) .. controls (128.95,360) and (120,342.09) .. (120,320) .. controls (120,297.91) and (128.95,280) .. (140,280) ;  
\draw  [draw opacity=0] (170,280) .. controls (170,280) and (170,280) .. (170,280) .. controls (164.08,280) and (158.76,285.15) .. (155.1,293.32) -- (170,320) -- cycle ; \draw   (170,280) .. controls (170,280) and (170,280) .. (170,280) .. controls (164.08,280) and (158.76,285.15) .. (155.1,293.32) ;  
\draw  [draw opacity=0] (154.98,346.42) .. controls (158.65,354.74) and (164.02,360) .. (170,360) -- (170,320) -- cycle ; \draw   (154.98,346.42) .. controls (158.65,354.74) and (164.02,360) .. (170,360) ;  
\draw  [draw opacity=0][dash pattern={on 0.84pt off 2.51pt}] (154.44,294.86) .. controls (151.66,301.73) and (150,310.48) .. (150,320) .. controls (150,329.53) and (151.67,338.28) .. (154.45,345.15) -- (170,320) -- cycle ; \draw  [dash pattern={on 0.84pt off 2.51pt}] (154.44,294.86) .. controls (151.66,301.73) and (150,310.48) .. (150,320) .. controls (150,329.53) and (151.67,338.28) .. (154.45,345.15) ;  
\draw  [draw opacity=0] (580,361) .. controls (580,361) and (580,361) .. (580,361) .. controls (513.73,361) and (460,343.09) .. (460,321) .. controls (460,298.91) and (513.73,281) .. (580,281) -- (580,321) -- cycle ; \draw   (580,361) .. controls (580,361) and (580,361) .. (580,361) .. controls (513.73,361) and (460,343.09) .. (460,321) .. controls (460,298.91) and (513.73,281) .. (580,281) ;  
\draw    (538.5,240.25) -- (538.5,217)(541.5,240.25) -- (541.5,217) ;
\draw [shift={(540,210)}, rotate = 90] [color={rgb, 255:red, 0; green, 0; blue, 0 }  ][line width=0.75]    (10.93,-4.9) .. controls (6.95,-2.3) and (3.31,-0.67) .. (0,0) .. controls (3.31,0.67) and (6.95,2.3) .. (10.93,4.9)   ;
\draw  [draw opacity=0][dash pattern={on 0.84pt off 2.51pt}] (140,160) .. controls (140,160) and (140,160) .. (140,160) .. controls (151.05,160) and (160,142.09) .. (160,120) .. controls (160,97.91) and (151.05,80) .. (140,80) -- (140,120) -- cycle ; \draw  [dash pattern={on 0.84pt off 2.51pt}] (140,160) .. controls (140,160) and (140,160) .. (140,160) .. controls (151.05,160) and (160,142.09) .. (160,120) .. controls (160,97.91) and (151.05,80) .. (140,80) ;  
\draw  [draw opacity=0][dash pattern={on 0.84pt off 2.51pt}] (470.84,113.12) .. controls (476.08,122.68) and (505.05,130) .. (540,130) .. controls (578.66,130) and (610,121.05) .. (610,110) -- (540,110) -- cycle ; \draw  [dash pattern={on 0.84pt off 2.51pt}] (470.84,113.12) .. controls (476.08,122.68) and (505.05,130) .. (540,130) .. controls (578.66,130) and (610,121.05) .. (610,110) ;  
\draw  [line width=3]  (460,319.5) .. controls (460,319.22) and (460.22,319) .. (460.5,319) .. controls (460.78,319) and (461,319.22) .. (461,319.5) .. controls (461,319.78) and (460.78,320) .. (460.5,320) .. controls (460.22,320) and (460,319.78) .. (460,319.5) -- cycle ;

\draw (317,112.4) node [anchor=north west][inner sep=0.75pt]  [color={rgb, 255:red, 208; green, 2; blue, 27 }  ,opacity=1 ]  {$\alpha $};
\draw (547,162.4) node [anchor=north west][inner sep=0.75pt]  [color={rgb, 255:red, 208; green, 2; blue, 27 }  ,opacity=1 ]  {$\alpha $};
\draw (141,42.4) node [anchor=north west][inner sep=0.75pt]    {$S^{1} \times _{q}$};
\draw (514,22.4) node [anchor=north west][inner sep=0.75pt]    {$S^{1} \times q$};
\draw (321,312.4) node [anchor=north west][inner sep=0.75pt]  [color={rgb, 255:red, 208; green, 2; blue, 27 }  ,opacity=1 ]  {$\alpha $};
\draw (117,371.4) node [anchor=north west][inner sep=0.75pt]  [font=\footnotesize]  {$\sum _{\beta }\ket{D_{\beta }}\bra{D_{\beta }} \ $};
\draw (31,369) node [anchor=north west][inner sep=0.75pt]  [font=\footnotesize]  {$\bra{HS^{2}}$};
\draw (276,369) node [anchor=north west][inner sep=0.75pt]  [font=\footnotesize]  {$\ket{\alpha }$};
\draw (384,312.4) node [anchor=north west][inner sep=0.75pt]  [font=\large]  {$=$};
\draw (504,383.4) node [anchor=north west][inner sep=0.75pt]  [font=\footnotesize]  {$\bra{HS^{2}}\ket{D_{\alpha }}$};
\draw (131,251.4) node [anchor=north west][inner sep=0.75pt]    {$S^{1} \times _{q}$};
\draw (514,249.4) node [anchor=north west][inner sep=0.75pt]    {$S^{1} \times _{q}$};
\draw (354,292.4) node [anchor=north west][inner sep=0.75pt]  [font=\footnotesize]  {$\bra{D_{\beta }}\ket{\alpha } =\delta _{\alpha \beta }$};
\draw (147,202.4) node [anchor=north west][inner sep=0.75pt]    {$||$};
\draw (607,312.4) node [anchor=north west][inner sep=0.75pt]  [color={rgb, 255:red, 208; green, 2; blue, 27 }  ,opacity=1 ]  {$D_{\alpha }$};
\draw (620,105) node [anchor=north west][inner sep=0.75pt]   {$\mathcal{Z}_{\alpha }$};

\end{tikzpicture}
    \caption{The first line shows the compactification of the cigar partition function to the $\mathbb{P}^1$ partition function based at $\alpha$. The second shows the equivalence of the cigar partition function and the twisted hemisphere partition function with an exceptional Dirichlet boundary condition. We denote the state generated on $T^2 = \partial (HS^2\times S^1)$ by the path integral for the topological twisted theory by $\bra{HS^2}$, and thus the hemisphere partition function by $\bra{HS^2}\ket{D_{\alpha}}$. The arrows $\Rightarrow$ indicates equality after removing the same, $\alpha$-dependent, perturbative contribution. }
    \label{fig:cigar_to_P1}
\end{figure}
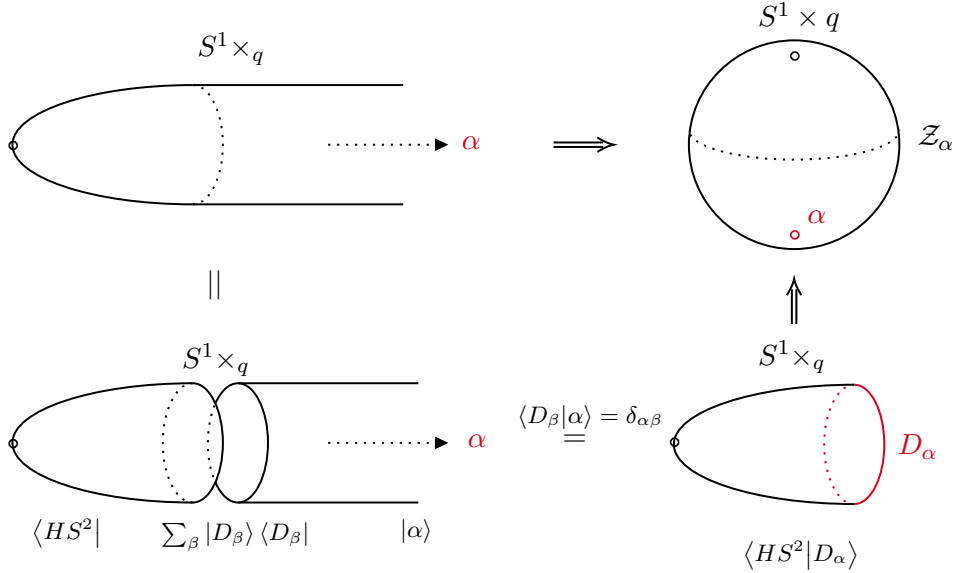

We begin with a seemingly different object; the topologically twisted partition function on the cigar in the presence of real masses $m \in \mathfrak{C}_H$, and a massive vacuum $\alpha$ at $\infty$. As the theory is topologically twisted, the cigar partition function may be regarded as the partition function on $\mathbb{R}^2 \times_q S^1$ with a massive vacuum $\alpha$ at the asymptotic boundary of $\mathbb{R}^2$, the BPS locus is unaffected. The cigar partition function should therefore equal, \textit{up to perturbative contributions},\footnote{That this equivalence holds only up to perturbative (classical and $1$-loop) contributions is familiar from four dimensions, where the Seiberg-Witten partition function is equivalent to the instanton partition function $S^4$ only up to the inclusion of perturbative contributions.} the partition function on $\mathbb{P}^1$ where we demand that the field configurations integrated over equal the vacuum $\alpha$ at the south pole. We denote this partition function $\mathcal{Z}_{\alpha}$. We will argue momentarily that the latter is precisely the vertex function $\mathsf{V}_{\alpha}$. These equivalences are illustrated in figure \ref{fig:cigar_to_P1}.

To complete the argument, it remains to show that the cigar partition function is equal to the twisted index on the hemisphere $\mathcal{Z}_{D_{\alpha}}$. We consider slicing the cigar path integral at some intermediate point in the infinitely long flat region where the twist becomes trivial and the theory is flat. This procedure is as follows.\footnote{Note that such cutting and gluing constructions have also been studied extensively in the literature and have been analysed systematically in \cite{dedushenko2020gluing,dedushenko2021gluing}.} In \cite{bullimore20223d}, it has been shown that the exceptional Dirichlet boundary conditions $\{D_{\alpha}\}$ generate a\textcolor{teal}{} complete basis of states $\ket{D_{\alpha}}$ on an elliptic curve $T^2$ for a 3d $\mathcal{N}=4$ theory. Let us work in a convention where we regard the boundary of $HS^2 \times S^1$ as a right boundary. Then, from \cite{bullimore20223d}, the dual states $\bra{D_{\alpha}}$ are similarly generated by left exceptional Dirichlet boundary conditions, defined in precisely the same way as for the right, but now supported on the \textit{repelling} holomorphic Lagrangian for the Morse flow generated by $m$. Alternatively, they are the same as the right exceptional Dirichlet boundary conditions for the opposite chamber of mass parameters $m \in \mathfrak{C}^{\text{opp}}_H$. For brevity, our notation does not distinguish between left and right boundary conditions.

Therefore, slicing the cigar path integral amounts to inserting the identity operator $\sum_{\beta} \ket{D_{\beta}} \bra{D_{\beta}} = \mathbb{I}$. It is also shown in \cite{bullimore20223d} that the path integral on the half space $T^2 \times \mathbb{R}^+$, with the left exceptional Dirichlet boundary condition at the origin of $\mathbb{R}^+$, and a massive vacuum $\alpha$ at $\infty$ is simply:
\begin{equation}
    \bra{D_{\beta}} \ket{\alpha} = \delta_{\alpha \beta},
\end{equation}
which follows by demonstrating that all overlaps on $T^2 \times I$ with the state $\ket{D_{\alpha}}$ agree with the boundary amplitudes computed on $T^2 \times \mathbb{R}^+$ with a vacuum $\alpha$ at $\infty$ (such overlaps transform identically over the space of fugacities in a way fixed by the effective Chern-Simons levels in a vacuum and boundary 't Hooft anomalies respectively). 

Combining these facts, we see that the partition function $\mathcal{Z}_{D_{\alpha}}$ computed in the previous section must agree with the cigar partition function (see figure \ref{fig:cigar_to_P1}) and thus, up to perturbative contributions, $\mathcal{Z}_{\alpha}=\mathsf{V}_\alpha$; the equivariant count of quasimaps based at $\alpha$. Physically, the perspective of these objects as twisted hemisphere partition functions is perhaps more natural, due to the role they play in holomorphic factorisation \cite{beem2014holomorphic}; their relation to counts of boundary operators via the state-operator correspondence \cite{bullimore2016boundaries,bullimore2021boundaries}; and the fact that they are more directly amenable to localisation. Geometrically however, it is arguably more natural to move to the perspective of quasimaps and the based $\mathbb{P}^1$ index, which we now turn to.

\subsection{Localisation on based \texorpdfstring{$\mathbb{P}^1$}{}}

In this section we compute the topologically twisted index for $\text{SQED}[N]$ on $\mathbb{P}^1$ with the following condition at the south pole: 
\begin{equation}
\begin{aligned}\label{eq:p1bc}
&\qquad\qquad\quad X_\alpha \neq 0,\\
&X_i = \bar{X}_i=\psi^{X_i} =\bar{\psi}^{X_i}= 0, \quad i\neq \alpha,\\
&\quad\,\, Y_i = \bar{Y}_i=\psi^{Y_i} =\bar{\psi}^{Y_i}= 0.
\end{aligned}
\end{equation}
These conditions are compatible with Higgs branch localisation. The vortex number is the degree $k$ of the $U(1)$ gauge bundle with $k = -\int_{HS^2} F_{12}/(2\pi)$. We write the $U(1)$ gauge bundle as $\mathcal{O}(k)$ to emphasise its holomorphic structure. We lift the $SO(2)_E$ rotations of $\mathbb{P}^1$ to the total space of the gauge bundle so that it fixes the fibre above the south pole. The BPS locus is determined by the same vortex equations as before
\begin{equation}\label{eq:p1bps}
\begin{aligned}
&\quad  2iF_{z\bar{z}} = \tau-2\mu_{\mathbb{R}},\quad\mu_{\mathbb{C}}
\vcentcolon= X\cdot Y = 0, \quad F_{z3} = 0, \\
&(\nabla_{\bar{z}} + iA_{\bar{z}})X = 0, \quad (D_3 + iA_3 + iA_3^t/2 + iA_3^i)X_i=0, \\
&(\nabla_{\bar{z}} - iA_{\bar{z}})Y = 0, \quad (D_3 -iA_3+iA_3^t/2 - iA_3^i)Y_i =0,\\
&\qquad\quad (\sigma + m_i + m_t/2)X_i = 0, \quad \phi_{\bar{z}}X_i = 0,\\ 
&\qquad\quad (-\sigma - m_i + m_t/2)Y_i = 0, \quad \phi_{\bar{z}}Y_i = 0,\\
&\qquad\qquad\qquad\qquad \nabla_\mu \sigma = \nabla_\mu \phi_{\bar{z}} = 0,\\
&\qquad\qquad\qquad\quad X_\alpha \neq 0 \text{ at south pole.}
\end{aligned}
\end{equation}
We note that $X_\alpha$ is covariantly holomorphic so its zeros at constant $x^3$ are isolated. Hence, when the real masses are generic and $X_\alpha\neq 0$, \eqref{eq:p1bps} implies $\sigma = -m_\alpha - m_t/2$ everywhere, $X_i = 0$ for  $i\neq \alpha$, $Y_i = 0$ for all $i$, and $\phi_{\bar{z}} = 0$.

The twisted boundary condition $X_\alpha(x^3 = \beta) = q^{-J}X_\alpha (x^3 = 0)$ implies that the only zero of $X_\alpha$ is at the north pole. The solution to these equations at each time slice exists and is unique (up to gauge transformation) provided $\tau$ is sufficiently large \cite{manton2023neumann}. We choose a gauge in which the spatial part of the gauge field is $SO(2)_E$ invariant and time independent. The condition $F_{z3}= 0$ implies that $A_3$ is constant along each time slice. We further gauge fix by demanding that $A_3$ is constant along the time direction. Since $X_\alpha\neq 0$ at the south pole and $X_\alpha(x^3 = \beta) = q^{-J}X_\alpha(x^3 = 0)$ then $X_\alpha(x^3 = \beta) = X_\alpha(x^3 = 0)$ (recall that $SO(2)_E$ acts trivially on the fibre above the south pole). The last equation in \eqref{eq:p1bps} implies $A_3 =- A_3^\alpha - A_3^t/2$ everywhere\footnote{Up to the addition of integer multiples of $\pi/\beta$ corresponding to gauge equivalent vortex configurations.} on $HS^2\times S^1$. In this gauge, the vortex equations can be written as
\begin{equation}
\begin{aligned}
& 2iF_{z\bar{z}} = \tau-\bar{X}_\alpha X_\alpha,\quad A_3 + A_3^t/2 + A_3^\alpha = 0,\\
&\qquad\quad(\nabla_{\bar{z}} + iA_{\bar{z}})X_\alpha = 0, \quad\nabla_3 X_\alpha = 0,\\
&\qquad\quad X_\alpha \neq 0 \text{ at the south pole},\\
&\qquad\quad\sigma =-m_\alpha - m_t/2,\quad \phi_{\bar{z}} = 0.\\
\end{aligned}
\end{equation}
We first study the unmatched modes which lie in the kernel/cokernel of $\nabla_{\bar{z}}$. We have
\begin{equation}
\begin{aligned}
&\delta X_i,\psi^{X_i} \in H^0(\mathcal{O}(k)), \quad \psi_{\bar{z}}^{X_i}\in H^1(\mathcal{O}(k)) = 0,\\
&\delta Y_i, \psi^{Y_i} \in H^0(\mathcal{O}(-k))=0, \quad \psi_{\bar{z}}^{Y_i}\in H^1(\mathcal{O}(-k)),
\end{aligned}
\end{equation}
hence only $\delta X_i, \psi^{X_i}, \psi_{\bar{z}}^{Y_i}$ have unmatched modes.
Let $s_{-k},...,s_0$ be a basis of holomorphic sections $H^0(\mathcal{O}(k))$, where $s_p$ has a zero of order $-p$ at the south pole. Similarly, let $u_1,...,u_{k-1}$ be a basis of $H^1(\mathcal{O}(-k))\cong H^0(O(k)\otimes \Omega^{1,0})^{\vee}$, where the dual basis of $u_p$ has a zero of order $p-1$ at the origin. We may then express the unmatched modes as
\begin{equation}
\begin{aligned}
X_i &= s_p \exp((2\pi in + p\epsilon)x^3/\beta),\qquad p = -1,...,-k\\
\psi^{X_i} &=  s_p \exp((2\pi in + p\epsilon )x^3/\beta), \qquad p = -1,...,-k\\
\psi^{Y_i}_{\bar{z}} &= u_p\exp((2\pi in + p\epsilon)x^3/\beta),\qquad p = 1,...,k-1
\end{aligned}
\end{equation}
The $p = 0$ case for $X_i,\psi^{X_i}$ is removed due to the boundary condition at the south pole.

Now we compute the angular momenta of the unmatched modes. The $SO(2)_E$ rotation of $\mathbb{P}^1$ acts trivially on the fibre of $\mathcal{O}(k)$ above the south pole, therefore $s_0$ has zero angular momentum. We choose a convention such that the angular momentum increases with the degree of zero at the north pole, and so the angular momentum of $s_p$ is $p$. Next, $SO(2)_E$ acts with weight one on the cotangent fibre above the south pole. So the dual of $u_p$ in $H^0(O(k)\otimes \Omega^{1,0})$ has angular momentum $-(p-1)-1 = -p$. the angular momentum of $u_p$ is $p$.

The contribution from $X_i$ and $\psi^{X_i}$ to the one-loop determinant is then
\begin{equation}
\begin{aligned}
&\det(D_3 - \sigma - m_i - m_t/2)^{-1} \\
&\qquad= \pm\prod_{\substack{n\in\mathbb{Z}\\ p=-1,...,-k}} (2\pi in + p\epsilon +i\beta( A_3+ A_3^i + A_3^t/2) -\beta(\sigma + m_i + m_t/2))^{-1}\\
&\qquad=\pm\prod_{p=-1,...,-k}\sinh(p\epsilon + i\beta(A_3^i - A_3^\alpha) -\beta (m_i- m_\alpha))^{-1},
\end{aligned}
\end{equation}
where we recall that $A_3 = -A_3^\alpha - A_3^t/2$ and $\sigma = -m_\alpha-m_t/2$ at the BPS locus. Notice that when $i = \alpha$ and $p = 0$ we get $\sinh(0)^{-1}$, this zero will be cancelled by the constant mode of the gaugino $\lambda$. A similar cancellation will be seen later in the construction of the virtual tangent bundle. 

The contribution from $\psi^{Y_i}_{\bar{z}}$ to the one-loop determinant is
\begin{equation}
\begin{aligned}
&\det(D_3 + \sigma + m_i-m_t/2) \\
&\qquad= \pm\prod_{\substack{n\in\mathbb{Z}\\ p=1,...,k-1}} (2\pi in + p\epsilon - i\beta ( A_3+ A_3^i - A_3^t/2 ) + \beta (\sigma +  m_i- m_t/2))\\
&\qquad=\pm\prod_{p=1,..,k-1}\sinh(p\epsilon +i\beta(A_3^\alpha - A_3^i + A_3^t) - \beta (m_\alpha - m_i + m_t)).
\end{aligned}
\end{equation}
Hence the one-loop determinant of all of the matter chirals in the $k$-vortex background is
\begin{align}
\prod_{i=1}^N\left(
\frac{(qtx_\alpha/x_i;q)_{k-1}}{(qx_\alpha/x_i;q)_k}\frac{\prod_{j=1}^k(x_\alpha^{1/2}x_i^{-1/2}q^{j/2})}{\prod_{j=1}^{k-1} (t^{1/2}x_\alpha^{1/2}x_i^{-1/2}q^{j/2})}\right).
\end{align}
The adjoint chiral contributes $t^{-1/2}(1 - t)$ to the one-loop determinant, this term arises from the constant mode of the fermionic scalar $\eta$.
Combining these ingredients, the based $\mathbb{P}^1$ index is then given by:\footnote{The sum at $k = 0$ involves a Pochhammer symbol of negative degree $(qt x_\alpha/x_i;q)_{-1} = (1 - t x_\alpha/x_i)^{-1}$. This (1-loop contribution) comes from the unmatched mode of $Y_i$ and $\psi^{Y_i}$.}\textsuperscript{,}\footnote{The $\hat{a}$-genus (acting on weights by $\hat{a}(x) = \frac{1}{x^{1/2}-x^{-1/2}}$ and satisfying $\hat{a}(x+y) = \hat{a}(x) \hat{a}(y)$) rather than the, perhaps more familiar, plethystic exponential appears because of the symmetrisation factor on the virtual structure sheaf \eqref{eq:virtualstructuresheaf}. The positive degree condition is related to the choice of stability condition, and therefore FI parameter sign.}
\begin{equation}\label{eq:p1result}
\begin{aligned}
\mathcal{Z}_\alpha(t,q,x_i,\xi) &= \frac{1-t}{t^{1/2}}
t^{N/2}\prod_{i}(x_\alpha^{1/2}x_i^{-1/2})\sum_{k \ge 0} (\xi q^{N/2}t^{-N/2})^k \prod_{i=1}^N\frac{(qtx_\alpha/x_i;q)_{k-1}}{(qx_\alpha/x_i;q)_k}\\
&= \prod_{i\neq \alpha}\frac{t^{1/2}x_\alpha^{1/2}x_i^{-1/2}}{1-tx_\alpha/x_i}\,\mathcal{Z}^{\text{vortex}}_{\alpha} \\
&= \hat{a}(P_{\alpha}) \, \mathcal{Z}^{\text{vortex}}_{\alpha} \,, 
\end{aligned}
\end{equation}
where $P_\alpha$ is the polarisation bundle at the fixed point $\alpha$. The vortex contribution matches the vortex contribution to the hemisphere partition function in equation \eqref{eq:vortex_pfn}, repeated here
\begin{equation}
    \mathcal{Z}^{\text{vortex}}_\alpha = \sum_{k\geq 0}\xi^k(qt^{-1})^{Nk/2}\prod_{i=1}^N \frac{(tx_\alpha x_i^{-1};q)_k}{(qx_\alpha x_i^{-1};q)_k} \,.
\end{equation}
We expect that this relation $\mathcal{Z}_\alpha =  \hat{a}(P_{\alpha}) \mathcal{Z}^{\text{vortex}}_{\alpha}$ should hold more generally for arbitrary theories, as the localisation computation proceeds similarly.

\subsection{Vortices and quasimaps}
The $\mathbb{P}^1$ degree $k$ BPS moduli space equations with masses switched off read
\begin{equation}\label{eq:p1bps_massless}
\begin{aligned}
&\qquad\qquad\qquad  2iF_{z\bar{z}} = \tau-2\mu_{\mathbb{R}}, \quad F_{z3} = 0, \\
&\qquad\qquad(\nabla_{\bar{z}} + iA_{\bar{z}})X = 0, \quad (D_3 + iA_3)X_i=0, \\
&\qquad\qquad(\nabla_{\bar{z}} - iA_{\bar{z}})Y = 0, \quad (D_3 -iA_3)Y_i =0,\\
&\qquad\quad \sigma \cdot X_i = 0, \quad \sigma \cdot Y_i = 0, \quad \phi_{\bar{z}}X_i = 0, \quad \phi_{\bar{z}}Y_i = 0\\ 
&\qquad \qquad\qquad\qquad \mu_{\mathbb{C}} \vcentcolon= X\cdot Y = 0, \\
&\qquad\qquad\qquad\quad X_\alpha \neq 0 \text{ at south pole.}
\end{aligned}
\end{equation}
These equations may be reframed (where the real moment map is traded for a stability condition and $\tau \to \infty$) as the complex algebraic data of a line bundle $\mathcal{O}(k)$ together with sections $(X_1,\ldots,X_N,Y_1,\ldots Y_N)$ satisfying the complex moment map $\mu_{\mathbb{C}} = 0$. This is precisely the data of the moduli space of quasimaps\footnote{Intuitively, at least for quiver gauge theory examples, the Higgs branch $\mathcal{M}_H$ carries tautological bundles $\mathcal{V}$ (associated to gauge symmetry) and $\mathcal{W}$ associated to flavour symmetry. A map $f: \mathbb{P}^1 \to \mathcal{M}_H$ induces pull-back bundles $f^*\mathcal{B}$ and $f^* \mathcal{W}$ on $\mathbb{P}^1$ and the quasimap picture works instead with this moduli space of bundles. We refer the reader to \cite{okounkov2015lectures} for a more thorough review of quasimap geometry.} $\textsf{QM}^{k} = \{ f: \, \mathbb{P}^1 \rightsquigarrow \mathcal{M}_H \}$ as we now explain. Switching the masses on, we will show that the equivariant Euler characteristic of this moduli space essentially coincides with the path integral localisation computation performed in the previous section.
For ease of notation we introduce a rank $N$ topologically trivial bundle $\mathcal{W} \to \mathbb{P}^1$ carrying a representation of $T_H$ with Chern character $\mathcal{W} = x_1 + \ldots + x_N$. We may then write our section as an element\footnote{With a slight abuse of notation conflating the quasimap $f$ and the bundle and section data.} $f$ of $t^{1/2} \cdot \mathcal{O}(k) \otimes \mathcal{W} \oplus t^{1/2} \cdot \mathcal{O}(-k) \otimes \mathcal{W}^{\vee}$
 
Our boundary condition choice \eqref{eq:p1bc} further implies that $f(0)$ lies in a fixed point $\alpha$ of the Higgs branch $\mathcal{M}_H^{T_H}$. A quasimap $f$ is in addition called non-singular at $p$ if $f(p)$ lies in the set of stable points of $\mathcal{M}_H$, and under our assumptions $f$ has at most finitely many singularities so that we recover the moduli space of \textit{based} quasimaps $\textsf{QM}^{k}_{\alpha}$.

\subsection{Vertex functions and the index}
The moduli space of quasimaps $\textsf{QM}_{\alpha}^{k}$ admits a natural torus action of $T = T_H \times \mathbb{C}^{\times}_q$  induced by the maximal torus of the Higgs branch flavour symmetry $T_H \subset G_H$ and the natural $\mathbb{C}_q^{\times}$ action on $\mathbb{P}^1$ which acts on sections by $z \to q z$. For the theories considered in the present work, this action has isolated fixed points. In particular, for the $\text{SQED}[N]$ example, writing $\omega$ for a torus weight of $T_H \times \mathbb{C}^{\times}_q$, we may write the fixed line bundle as $\omega \cdot \mathcal{O}(k)$ and the only non-vanishing section at $\infty$ is $z^{k}$ so that for a fixed point $\alpha=1,\ldots, N$ it follows $\omega = q^{-k}x_i$ so that the fixed sections are 
\begin{equation}
    \mathcal{V} = x_\alpha  \cdot \mathcal{O}(k) \,, \quad \mathcal{W} = x_1 + \ldots + x_N \,.
\end{equation}

The moduli space $\textsf{QM}_{\alpha}^{k}$ admits a perfect deformation-obstruction theory that allows us to perform localisation computations \cite{ciocan2014stable}. We will now proceed to identify the $\mathbb{P}^1$ localisation result \eqref{eq:p1result} with the so-called vertex function $\mathsf{V}_{\alpha}$: the generating function of equivariant Euler characteristic of the virtual structure sheaf on $\textsf{QM}_{\alpha}^{k}$.

We review the quasimap moduli space localisation computation briefly and refer the reader to \textit{e.g.} \cite{ciocan2014stable,okounkov2015lectures} for more details. The virtual tangent bundle encoding fluctuations above a fixed point at a point\footnote{Recall here that a point is a collection of bundles on $\mathbb{P}^1$ and a section $f$ satisfying $\mu_{\mathbb{C}} = 0$.} is given by
\begin{equation}
\label{eq:virtualtangent}
    T^{\text{vir.}} \textsf{QM}^{k}_{\alpha} = H^{\bullet}(T\mathscr{M}_H) - t^{1/2}\cdot P_{\alpha}^{\vee} \,,
\end{equation}
where $T\mathscr{M}_H$ is a bundle on $\mathbb{P}^1$ induced by the tangent bundle $T\mathcal{M}_H$ in equation \eqref{eq:sqedtangent}, specifically
\begin{equation}
    T\mathscr{M}_H = t^{1/2} \cdot \mathcal{W}^{\vee} \otimes \mathcal{V} - t^{1/2} \cdot \mathcal{V} \otimes \mathcal{V}^{\vee} + t^{1/2} \cdot \mathcal{W}\otimes \mathcal{V}^{\vee} - t^{1/2} \mathcal{V} \otimes \mathcal{V}^{\vee} \,.
\end{equation}
We import this construction from the geometry literature but we expect that physically the virtual tangent bundle encodes the bosonic and fermionic fluctuations about the vortex BPS equations \eqref{eq:p1bps_massless} on $\mathbb{P}^1$, analogously to the unbased case studied by Bullimore \textit{et. al.} \cite{bullimore2019twisted}. This identification will also be physically justified \textit{a posteriori} by matching the Higgs and Coulomb branch localisation calculations of section \ref{sec:localisation} to the Euler characteristic calculation to follow.

The symmetrised virtual structure sheaf corresponding to the virtual tangent bundle \eqref{eq:virtualtangent} is denoted
\begin{equation}\label{eq:virtualstructuresheaf}
    \hat{\mathcal{O}}_{\text{vir.}} = \mathcal{O}_{\text{vir.}} \otimes \mathcal{K}^{1/2}_{\text{vir.}} ,
\end{equation}
where $\mathcal{K}_{\text{vir.}} = \text{det} \,T^{\text{vir.}}\textsf{QM}^{k}_{\alpha}$ is the virtual canonical bundle and $\mathcal{O}_{\text{vir.}}$ is the virtual structure sheaf. The vertex function $\mathsf{V}_{\alpha}$ is then defined as the generating function of the (equivariant) Euler characteristics of the virtual structure sheaf over the quasimap moduli space degrees
\begin{equation}
    \mathsf{V}_{\alpha} = \sum_{k} \xi^{k} \chi_{\mathsf{T}}(\textsf{QM}^{k}_{\alpha}, \hat{\mathcal{O}}_{\text{vir.}}) \,.
\end{equation}
The vertex function may then be computed by localisation \cite{ciocan2014stable, graber1997localization}
\begin{equation}
\label{eq:vertexfunction}
    \mathsf{V}_{\alpha} = \hat{a}(P_{\alpha})\sum_{k>0} \sum_{f \in \left( \textsf{QM}^{k}_{\alpha}\right)^{\mathsf{T}} } \xi^{k} \hat{a}\left( H^{\bullet}(T\mathscr{M}_H) \right) \,.
\end{equation}
Counting these sections we find precisely the vortex contributions to the localisation formula \eqref{eq:localisationsummary}. In particular we need only note for example
\begin{equation}
    \text{Ch}_{T} x_i H^0(\mathcal{O}(k)) = x_i(1 + q + \ldots + q^{k}),   
\end{equation}
and note that for $k>0$ only the zeroth cohomology contributes to the virtual tangent bundle \eqref{eq:virtualtangent}. In terms of $q$-Pochhammer symbols we thus have
\begin{equation}
    \hat{a}\left(  H^{\bullet}(T\mathscr{M}_H) \right) = \left(q^{1/2}t^{-1/2}\right)^{N k}\prod_{i=1}^N \frac{(t x_i / x_{\alpha};q)_{k}}{(q x_i / x_{\alpha};q)_{k}} .
\end{equation}
Substituting into \eqref{eq:vertexfunction} we see that the normalised vertex function $\mathsf{V}_{\alpha}$ thus matches the based $\mathbb{P}^1$ result \eqref{eq:p1result}:
\begin{equation}
    \mathcal{Z}_{\alpha} = \mathsf{V}_{\alpha} .
\end{equation}

\subsection{Higgs and Coulomb branch localisation}\label{sec:handclocalisation}
In this section we discuss the geometrical interpretation of the Higgs and Coulomb branch localisation schemes discussed in section \ref{sec:localisation}. In particular, we argue that the Coulomb branch localisation scheme gives the Mellin-Barnes type $q$-Jackson integral formulation of vertex functions as in \cite{aganagic2017quasimap} whereas the Higgs branch scheme more naturally gives the equivariant Euler character localisation formula of equation \eqref{eq:vertexfunction}. 

In \cite{aganagic2017quasimap} it is argued that one may trade a descendant condition for a relative condition on the quasimap moduli space and that this explains the existence of certain $q$-Jackson integral formulae for vertex functions that are often used in the enumerative geometry literature. The result is an `off-shell' or `integrand' expression for the vertex of the schematic form
\begin{equation}\label{eq:offshell}
    \mathsf{V} = e^{\phi_0} \hat{a}\left[\frac{1-t^{-1}q}{1-q} T\mathcal{M}_H\right].
\end{equation}
The different vertex functions for each vacuum $\alpha$ are then given by a formal $q$-Jackson integral corresponding to evaluation at Chern roots (as in equation \eqref{eq:fugacity_subsitution}) shifted by powers of $q$, schematically
\begin{equation}
    \mathsf{V}_{\alpha} = \sum_{k \in \mathbb{Z}} \mathsf{V} \rvert_{s=s_\alpha q^k} \,.
\end{equation}
We provide a physical interpretation of this heuristic formula: it is exactly the Coulomb branch localisation procedure 
\eqref{eq:CBresult}. This physical construction has a number of interesting features. Most importantly, the physical localisation construction gives rise to a normalisation of the vertex (see equation \eqref{eq:localisationsummary}) that gives rise to solutions to $q$-difference equations\footnote{See \cite{dinkins2022exotic} for an excellent exposition of the relationships between $q$-difference equations and vertex functions.} and exactly factorises more general three-manifold partition functions \cite{bullimore2021boundaries}.

Let us now discuss the Coulomb branch localisation formula \eqref{eq:CBresult} from section \ref{sec:localisation}. Recall also, from section \ref{subsec:ed}, that the weights of the hypermultiplet representation may be split into positive and negative weight spaces according to a chamber $\mathfrak{C}_H$ choice as
\begin{equation}
    T^*R = Q^{+} +  Q^{0} + \bar{Q}^{0} + Q^{-}.
\end{equation}
We may now write the $\text{SQED}[N]$ localisation formula \eqref{eq:CBresult} in terms of these weight space decompositions in a way that generalises naturally to other 3d $\mathcal{N}=4$ gauge theories. We have\footnote{The plethystic exponential (PE) here is defined formally on torus weights by the properties $\text{PE}[\omega] = \frac{1}{1-\omega}$, $\text{PE}[\omega + \omega'] = \text{PE}[\omega ]\text{PE}[\omega']$ and $\text{PE}[-\omega] = 1-\omega$.}
\begin{equation}\label{eq:generalised-localisation}
    \mathcal{Z}_{D_\alpha}(x_i,\xi, t,q) = \int_{0}^{s(\alpha)} d_q s \, e^{-\phi_0(s,\xi)} \hat{a} \left[ \frac{1-t^{-1}q}{1-q} \left( Q^{+} + \bar{Q}^{0} - \mu_{\mathbb{C}
    } \right) \right] ,
\end{equation}
where, for ease of notation, we conflate the weight space $\mu_{\mathbb{C}}$ and its character. The exponential prefactor is\footnote{In terms of Higgs branch geometry, $\phi_0(s,\xi)$ is the natural quadratic pairing between K\"ahler parameters (resolution parameters of the Higgs branch in $H^2(\mathcal{M}_H,\mathbb{Z})$) and the Chern roots of tautological line bundles (that generate $H^2(\mathcal{M}_H,\mathbb{Z})$ by Kirwan surjectivity \cite{mcgerty2018kirwan}).} 
\begin{equation}
    \phi_0(s,\xi) = \frac{\log \xi \log s}{\log q} \,.
\end{equation}        
The integral above is a $q$-Jackson integral\footnote{Precisely $\int_0^{a} d_q x f(x) = \sum_{n \in \mathbb{Z}}f( a q^n )$.} that evaluates the integrand at $q$-shifts of the Chern root evaluation $s_\alpha$.

We note here that the zeta function regularisations used in the localisation computations of section \ref{sec:localisation} may be re-expressed as regularising the $\hat{a}$ genus of weights as follows. Let us recall first the relationship between the $\hat{a}$-genus and the plethystic exponential
\begin{equation}
    \hat{a}(\omega) = \frac{1}{\omega^{-1/2} - \omega^{1/2}}  = \omega^{1/2} \text{PE}[\omega] \,,
\end{equation}
where $\omega$ is a torus weight (we assume $\omega>0$). The zeta regularised expression is then
\begin{equation}
\begin{split}
    \hat{a}\left(\frac{\omega}{1-q}\right) &= \prod_{n=0}^{\infty}(q^n \omega)^{1/2} \text{PE}\left[ \frac{w}{1-q}\right] \\
    &= \exp(\frac{1}{4} \log \omega - \frac{1}{4}\frac{\log(\omega)^2}{\log(q)}) \text{PE}\left[ \frac{w}{1-q}\right] ,
\end{split}    
\end{equation}
which simplifies further for $\mathcal{N}=4$ weight combinations to
\begin{equation}
    \hat{a}\left( \frac{1-t^{-1}q}{1-q} \omega \right) = \exp(-\frac{1}{4}\log(t^{-1}q) + \frac{\log(t^{-1}q)^2}{4 \log q} + \frac{\log \omega \log (t^{-1}q)}{2\log q}) \text{PE}\left[ \frac{1-t^{-1}q}{1-q} \omega \right].
\end{equation}
We may now alternatively write the weight space formulation of the Coulomb branch localisation formula \eqref{eq:generalised-localisation} in terms of the plethystic exponential as
\begin{equation}
    \mathcal{Z}_{D_\alpha}(x_i,\xi, t,q) = \int_{0}^{s(\alpha)} d_q s \, e^{-\phi(s,\xi)} \text{PE} \left[ \frac{1-t^{-1}q}{1-q} \left( Q^{+} + \bar{Q}^{0} - \mu_{\mathbb{C}} \right) \right] ,
\end{equation}
where\footnote{The term $-\mu_{\mathbb{C}}$ needs to be interpreted appropriately, for SQED$[N]$ it  gives $\log(q/t)/4 + (\log^2(t) - \log^2(q))/(4\log(q))$.}
\begin{equation}
\begin{aligned}
    \phi(s,\xi) = \phi_0(s,\xi) + \sum_{\omega \in Q^{+} + \bar{Q}^{0} - \mu_{\mathbb{C}}} -\frac{1}{4}\log(t^{-1}q) + \frac{\log(t^{-1}q)^2}{4 \log q} + \frac{\log \omega \log (t^{-1}q)}{2\log q}.
    \end{aligned}
\end{equation}
This expression is a general geometric proposal for the formula we expect from a careful Coulomb branch localisation in general for gauge theories satisfying our assumptions. We note that this expression also agrees with the anomaly polynomial calculations of \cite{bullimore2016boundaries} and the geometric prescription elucidated in \cite{Crew:2021ipc}.

We now consider evaluating the $q$-Jackson integral \eqref{eq:generalised-localisation}. It is straightforward to verify that evaluating the integral \eqref{eq:generalised-localisation} gives
\begin{equation}
    \mathcal{Z}_{D_\alpha}(x_i,\xi,t,q) = e^{\phi_{\alpha}}\mathcal{Z}^{\text{1-loop}}_{\alpha}\mathcal{Z}_{\alpha}^{\text{vortex}}(x,\xi,q,t) \,,
\end{equation}
where
\begin{equation}
    \phi_\alpha = \sum_{\omega \in T^+_{\alpha} \mathcal{M_H}}  -\frac{1}{2}\log(t^{-1}q) + \frac{\log(t^{-1}q)^2}{2 \log q} +  \frac{\log{\omega}\log t}{\log q},
\end{equation}
and, recalling from the discussion in section \ref{subsec:ed} that evaluating the bundle $Q^+ + \bar{Q}^0 - \mu_{\mathbb{C}}$ at a fixed point $s_\alpha$ gives $T_{\alpha}^+ \mathcal{M}_H$, the one-loop contribution (the $k=0$ sector) may be written geometrically as
\begin{equation}
    \mathcal{Z}^{\text{1-loop}}_{\alpha} = \text{PE}\left[ \frac{1-t^{-1}q}{1-q} T^{+}_{\alpha}\mathcal{M}_H \right]\,,
\end{equation}
which agrees with \eqref{eq:sqed-1loop} for $\text{SQED}[N]$.

In contrast, the Higgs branch localisation scheme does not involve a free gauge fugacity and arises after specialising the `off-shell' Coulomb branch localisation formula (the integrand of \eqref{eq:generalised-localisation}). The Higgs branch localisation thus more directly yields the equivariant Euler character formula \eqref{eq:vertexfunction} together with the same perturbative prefactor as the Coulomb branch expression above. 

\subsection{Pole subtraction and mirror symmetry}

With the geometric interpretation of the hemisphere partition function in hand, we now give a brief physical argument for the pole subtraction formula of Aganagic and Okounkov \cite{Aganagic:2016jmx, Aganagic:2017smx}.

In \cite{bullimore2021boundaries}, the mirror dual boundary condition to exceptional Dirichlet $D_{\alpha}$ was found to be a Neumann (for the gauge symmetry) type boundary condition, coupled to boundary $\mathbb{C}^\times$ valued matter. These are called \textit{enriched} Neumann boundary conditions and denoted by $\mathcal{N}_{\alpha}$. Let $\mathcal{T}$ be an $\mathcal{N}=4$ theory with Higgs branch $\mathcal{M}_H$ and Coulomb branch $\mathcal{M}_C$, with isolated massive vacua with generic mass parameters $m$ and FI parameters $\zeta$ are turned on. Its mirror $\widetilde{T}$ has Higgs and Coulomb branches switched, and we may label the vacua in the same way. The mass and FI parameters, as well as the chambers $\mathfrak{C}_H$ and $\mathfrak{C}_C$ are interpreted as the resolution and equivariant parameters/chambers for $\mathcal{M}_C$.

By mirror symmetry:
\begin{equation}
    \widetilde{\mathcal{Z}}_{\tilde{D}_{\alpha}}(t^{-1}q,q,\zeta,x_i) = \mathcal{Z}_{N_{\alpha}}(t,q,x_i, \zeta).
\end{equation}
The partition function on the left is the $U(1)_C$ twisted partition function for $\widetilde{\mathcal{T}}$, with an exceptional Dirichlet boundary condition $\tilde{D}_{\alpha}$ for the vacuum $\alpha$ in chamber $\mathfrak{C}_C$. The replacement $t \mapsto t^{-1}q$ arises due to the exchange of $R$-symmetries under mirror symmetry, and shifting between the $U(1)_H$ and $U(1)_C$ twist \cite{bullimore2021boundaries}. By the above arguments, it is proportional to the vertex function equivariantly counting quasimaps to $\mathcal{M}_H$, which we denote $\widetilde{\mathsf{V}}_{\beta}(\xi,x,q,t^{-1}q)$.

On the other hand, stretching the $\mathcal{Z}_{N_{\alpha}}$ partition function into a cigar, and inserting the identity $\sum_{\beta} \ket{D_{\beta}} \bra{D_{\beta}} = \mathbb{I}$, we have:
\begin{equation}
    \mathcal{Z}_{N_{\alpha}} = \bra{HS^2}\ket{N_{\alpha}} = \sum_{\beta} \bra{HS^2}\ket{D_{\beta}} \bra{D_{\beta}}\ket{N_{\alpha}},
\end{equation}
where we have used the same notation as at the beginning of section \ref{sec:quasimap}. We recognise $\bra{HS^2}\ket{D_{\beta}} $ as simply $\mathcal{Z}_{D_{\beta}}$, the twisted hemisphere index with exceptional Dirichlet boundary condition. The overlap $\bra{D_{\beta}}\ket{N_{\alpha}}$ is given by the path integral on $I \times T^2$, where $I$ is an interval (its length is $Q$-exact), with boundary conditions $D_{\beta}$ and $N_{\alpha}$ on either end. In \cite{bullimore20223d}, these were referred to as boundary amplitudes, and shown to be related to the elliptic stable envelopes of Aganagic and Okounkov \cite{Aganagic:2016jmx}:
\begin{equation}
    \bra{D_{\beta}}\ket{N_{\alpha}} = \frac{\text{Stab}(\alpha)_{\mathfrak{C}_H, \zeta}|_{\beta}}{\Theta(T^{+}_{\beta}\mathcal{M}_H )},
\end{equation}
where we have used the shorthand that if $W$ is a weight space, $\Theta(W) \vcentcolon= \prod_{\omega \in W} \vartheta(\omega)$, and $\vartheta$ is the Jacobi theta function.

Together, we obtain:\footnote{Proofs of such mirror symmetry relations for quiver variety examples may be found in the geometry literature \cite{dinkins2020symplectic, dinkins20223d}.}
\begin{equation}
    \widetilde{\mathcal{Z}}_{\tilde{D}_{\alpha}}(t^{-1}q,q,\zeta,x_i)  = \sum_{\beta} \frac{\text{Stab}(\alpha)_{\mathfrak{C}_H, \zeta}|_{\beta}}{\Theta(T^{+}_{\beta}\mathcal{M}_H )}\mathcal{Z}_{D_{\beta}}(t,q,x_i, \zeta).
\end{equation}
As shown in section \ref{sec:handclocalisation}, the twisted hemisphere indices $\widetilde{\mathcal{Z}}_{\tilde{D}_{\alpha}}$ and $\mathcal{Z}_{N_{\alpha}}$ are, up to prefactor, vertex functions $\widetilde{\mathsf{V}}_{\alpha}$ and $\mathsf{V}_{\alpha}$ counting quasimaps into $\mathcal{M}_C$ and $\mathcal{M}_H$ respectively. We thus recover precisely the pole subtraction formula of Aganagic and Okounkov. Note in our choice of chamber, $\text{Stab}(\alpha)_{\mathfrak{C}_H, \zeta}|_{\beta}$ is upper triangular in $\alpha$ and $\beta$. See §5 of \cite{Aganagic:2017smx} for mathematical details, in particular  §5.5 for the example in the case of self-mirror-dual SQED[$2$].

\acknowledgments

It is a pleasure to thank Mathew Bullimore, Adam Chalabi, Cyril Closset, Hunter Dinkins, Nick Dorey, Andrea Ferrari, Mark Gross, Nick Manton, Ivan Smith, Michael Walter, Claude Warnick, and Yutaka Yoshida for helpful discussions. SC and DZ are grateful to ICMAT for their hospitality while part of this work was completed. BZ is supported by a Trinity College internal graduate studentship. DZ is supported by a Junior Research Fellowship from St. John’s College, Oxford.

\appendix

\section{Lagrangians and boundary terms}\label{ap:lagrangians}

In this section we give more details on the twisted 3d $\mathcal{N}=2$ Lagrangians. We also carefully argue that the boundary terms do not contribute to the path integral. The vector multiplet Lagrangian is $Q$-exact:
\begin{equation}
\epsilon\bar{\epsilon} \mathcal{L}_{\text{vec}} = \delta_{\epsilon}\delta_{\bar{\epsilon}}(\lambda\bar{\lambda} -\lambda_{\bar{z}}\bar{\lambda}_z - 4\sigma D).
\end{equation}
It can be written as
\begin{equation}
\mathcal{L}_{\text{vec}} =  \mathcal{L}_{\text{vec}}^{B}+  \mathcal{L}_{\text{vec}}^{F},
\end{equation}
where the bosonic and fermionic parts are given by:
\begin{equation}\label{eq:appendix_bosonic_lagrangian}
\begin{aligned}
 \mathcal{L}_{\text{vec}}^{B} 
 =&\,(\delta_\epsilon \lambda\delta_{\bar{\epsilon}}\bar{\lambda} - \delta_{\epsilon}\lambda_{\bar{z}}\delta_{\bar{\epsilon}}\bar{\lambda}_z - 4\delta_\epsilon \sigma \delta_{\bar{\epsilon}}D - 4\delta_{\bar{\epsilon}}\sigma\delta_\epsilon D)/(\epsilon\bar{\epsilon})\\
 =&\,(2iF_{z\bar{z}} - \nabla_3\sigma + D)(2iF_{z\bar{z}}+ \nabla_3\sigma + D) +4 (F_{z3} + i\nabla_z\sigma)(F_{\bar{z}3} + i\nabla_{\bar{z}}\sigma)\\
&+4\sigma(-2i\nabla_z(-F_{\bar{z}3} - i\nabla_{\bar{z}}\sigma) - \frac{1}{2}\nabla_3(-2iF_{z\bar{z}} + \nabla_3\sigma - D))\\
&+2(-2iF_{z\bar{z}} + \nabla_3\sigma - D)D\\ 
=&\,4|F_{z\bar{z}}|^2 + 4F_{z3}F_{\bar{z}3} + \nabla_\mu\sigma\nabla^\mu\sigma -D^2,\\
\end{aligned}
\end{equation}
\begin{equation}
\begin{aligned}
\mathcal{L}_{\text{vec}}^{F} =&\, (\lambda \delta_\epsilon\delta_{\bar{\epsilon}} \bar{\lambda} -\lambda_{\bar{z}}\delta_{\epsilon}\delta_{\bar{\epsilon}}\bar{\lambda}_z - 4\delta_{\epsilon} \sigma \delta_{\bar{\epsilon}}D - 4\delta_{\bar{\epsilon}}\sigma\delta_{\epsilon}D)/(\epsilon\bar{\epsilon})\\
=&\, \lambda(\nabla_{\bar{z}}\bar{\lambda}_z - \frac{1}{2}\nabla_3\bar{\lambda} - \nabla_{\bar{z}}\bar{\lambda}_z - \frac{1}{2}\nabla_3\bar{\lambda})
+\lambda_{\bar{z}}(-\nabla_z \bar{\lambda} +\nabla_3\bar{\lambda}_z) +\nabla_z \bar{\lambda}\\
&+2\bar{\lambda}(-\nabla_z\lambda_{\bar{z}} + \nabla_3\lambda/2) +2\lambda(\nabla_{\bar{z}}\bar{\lambda}_z + \nabla_3\bar{\lambda}/2)\\
=&\,\bar{\lambda}\nabla_3\lambda +\lambda_{\bar{z}}\nabla_3\bar{\lambda}_z+2\lambda\nabla_{\bar{z}}\bar{\lambda}_z-2\bar{\lambda}\nabla_z\lambda_{\bar{z}}\\
=&\,\frac{1}{2}\Big(\bar{\lambda}\nabla_3\lambda +\lambda\nabla_3\bar{\lambda}+\lambda_{\bar{z}}\nabla_3\bar{\lambda}_z+\bar{\lambda}_z\nabla_3\lambda_{\bar{z}} +2\lambda\nabla_{\bar{z}}\bar{\lambda}_z\\
&\qquad + 2\bar{\lambda}_z\nabla_{\bar{z}}\lambda-2\bar{\lambda}\nabla_z\lambda_{\bar{z}}-2\lambda_{\bar{z}}\nabla_z\bar{\lambda}\Big).
\end{aligned}
\end{equation}

To derive the expression for the bosonic Lagrangian in the last line of \eqref{eq:appendix_bosonic_lagrangian}, one must integrate by parts the term $4i\nabla_z\sigma F_{\bar{z}3} + 4i\nabla_z\sigma F_{\bar{z}3} -8\sigma\nabla_z\nabla_{\bar{z}}\sigma$. The boundary term $\int_{\partial} 2i\sigma F_{n3} -2\sigma\nabla_n\sigma$ is the $Q$-variation of $-\int_\partial \sigma\lambda_n$ so can be discarded. In passing to the last line of the fermionic Lagrangian, the boundary terms arising from integration by parts also cancel since
\begin{equation}
\begin{aligned}
\int_{HS^2} \nabla_{\bar{z}}(\lambda\bar{\lambda}_z)+\nabla_z(\lambda_{\bar{z}}\bar{\lambda})
&= \frac{1}{4} \int_{HS^2} \nabla\cdot ((\lambda - \bar{\lambda})\hat{\lambda}) -i \nabla \times ((\lambda + \bar{\lambda})\hat{\lambda}) \\
&= \frac{1}{4} \int_{\partial HS^2} (\lambda - \bar{\lambda})\hat{\lambda}_n -i (\lambda + \bar{\lambda})\hat{\lambda}_t  = 0,
\end{aligned}
\end{equation}
where we write $\hat{\lambda} = \bar{\lambda}_zdz + \lambda_{\bar{z}}d\bar{z}$. We may also drop derivatives in the $x^3$-direction by integrating by parts.

The chiral multiplet Lagrangian may be written in the $Q$-exact form:
\begin{equation}
\begin{aligned}
\epsilon\bar{\epsilon}\mathcal{L}_{\text{chi}} &= \delta_{\epsilon}\delta_{\bar{\epsilon}}(-\bar{\psi}\psi +\bar{\psi}_z\psi_{\bar{z}}+2\bar{X}\sigma X),
\end{aligned}
\end{equation}
and split into
\begin{equation}
\mathcal{L}_{\text{chi}} = \mathcal{L}_{\text{chi}}^{B} + \mathcal{L}_{\text{chi}}^{F},
\end{equation}
where the bosonic and fermionic parts are given by:
\begin{equation}
\begin{aligned}
\mathcal{L}_{\text{chi}}^{B}
=\,&(-\delta_{\bar{\epsilon}}\bar{\psi}\delta_{\epsilon}\psi + \delta_{\bar{\epsilon}}\bar{\psi}_z\delta_{\epsilon}\psi_{\bar{z}} + \delta_\epsilon\bar{\psi}_z\delta_{\bar{\epsilon}}\psi_{\bar{z}} + 2\bar{X}\delta_{\epsilon}\delta_{\bar{\epsilon}}\sigma X + 2\bar{X}\sigma \delta_{\epsilon}\delta_{\bar{\epsilon}}X)/(\epsilon\bar{\epsilon})\\
=\,& \,(\nabla_3 - iA_3 + \sigma)\bar{X}(\nabla_3 + iA_3 - \sigma)X + 4(\nabla_z - iA_z)\bar{X}(\nabla_{\bar{z}} + iA_{\bar{z}})X
\\&-F_{\bar{z}}\bar{F}_z+\bar{X}(2iF_{z\bar{z}} - D_3\sigma + D)X - 2\bar{X}\sigma (\nabla_3 + iA_3 - \sigma)X\\
=\,&\, (\nabla_3 - iA_3)\bar{X}(\nabla_3 + iA_3)X + 4(\nabla_z - iA_z)\bar{X}(\nabla_{\bar{z}} + iA_{\bar{z}})X \\
&+ \sigma^2|X|^2 + D|X|^2 +2iF_{z\bar{z}}|X|^2 - F_{\bar{z}}\bar{F}_z,\\
\mathcal{L}_{\text{chi}}^{F} =& (-\delta_\epsilon\delta_{\bar{\epsilon}}\bar{\psi}\psi + \delta_{\epsilon}\delta_{\bar{\epsilon}}\bar{\psi}_z\psi_{\bar{z}} +\bar{\psi}_z\delta_{\epsilon}\delta_{\bar{\epsilon}}\psi_{\bar{z}} + 2\delta_{\epsilon}\bar{X}\delta_{\bar{\epsilon}}\sigma X + 2\delta_{\epsilon}\bar{X}\sigma\delta_{\bar{\epsilon}}X + 2\bar{X}\delta_{\epsilon}\sigma
\delta_{\bar{\epsilon}}X)/(\epsilon\bar{\epsilon})\\
 =\,&-\bar{\psi}(\nabla_3 + iA_3 + \sigma)\psi - \bar{\psi}_z(\nabla_3 + iA_3 - \sigma)\psi_{\bar{z}} + 2\bar{\psi}(\nabla_z + iA_z)\psi_{\bar{z}}  \\ &-2\bar{\psi}_z(\nabla_{\bar{z} } + iA_{\bar{z}})\psi +\bar{\lambda}_{z}\bar{X}\psi_{\bar{z}} +\bar{\psi}_z\lambda_{\bar{z}}X + \bar{\psi}\lambda X - \bar{X}\bar{\lambda}\psi.
\end{aligned}
\end{equation}

\section{Eigenvalue problem for the Neumann boundary condition}\label{ap:eigenvalue_neumann}
In this appendix we argue that the hemisphere Laplacian with a Neumann boundary condition admits an orthonormal basis of eigenfunctions. We consider the 2d hemisphere Laplacian $-4D_zD_{\bar{z}}$ acting on functions with a Neumann boundary condition $D_{\bar{z}}f|_\partial = 0$, where the gauge covariant derivative $D_i = \nabla_i + iA_i$ may include a background $U(1)$ connection. We will show that $-4D_zD_{\bar{z}}$ is Hermitian with respect to the standard $L^2$ inner product $\int_{HS^2}\bar{f}g$ on the space of functions satisfying the Neumann boundary condition $D_{\bar{z}}f|_\partial = D_{\bar{z}}g|_\partial = 0$. As a result, eigenfunctions with different eigenvalues are then orthogonal. 

To prove hermiticity of $-4D_zD_{\bar{z}}$, it is equivalent to work with the Laplacian $-D_iD^i$ since this differs from $-4D_zD_{\bar{z}}$ by a real-valued function (a Hermitian operator). By Stokes' theorem we have
\begin{equation}
\int_{HS^2} \bar{g}D_iD^i f - D_iD^i\bar{g}f  = \int_{\partial HS^2}(\bar{g}D_nf - D_n\bar{g}f),
\end{equation}
where $D_t$ and $D_n$ denote the boundary gauge-covariant tangential and normal derivative respectively.
Using $D_nf = iD_tf$ and $D_ng = iD_tg$, the equation above may be rewritten as:
\begin{equation}
\int_{\partial HS^2}(\bar{g}D_nf - D_n\bar{g}f) = i\int_{\partial HS^2}\bar{g}D_tf + D_t\bar{g}f = i\int_{\partial HS^2}D_t(\bar{g}f) = 0.
\end{equation}
Therefore the Laplacian is Hermitian and the eigenfunctions are orthogonal.

Now we show that the eigenfunctions are complete. Firstly, we choose an orthonormal basis of $(0,1)$ form eigenfunctions $\{f^i\}_{i \in \mathbb{Z}_{\ge 0}}$ that satisfy the Dirichlet boundary condition
\begin{equation}
-4D_{\bar{z}}D_zf^i_{\bar{z}} = \lambda_i f^i_{\bar{z}}, \qquad f^i_{\bar{z}} = 0\text{ on } \partial HS^2, \quad i \in \mathbb{Z}_{\ge 0},
\end{equation}
where $f^i_{\bar{z}}$ denotes the component of $f^i$ in the local orthonormal frame component $\nabla_{\bar{z}}$.
The operator $-4D_{\bar{z}}D_z$ has a trivial kernel since any function $f_{\bar{z}}^i$ that satisfies $D_zf_{\bar{z}}^i = 0$ is identically zero due to the boundary condition $D_{\bar{z}}f^i_{\bar{z}}|_\partial = 0$. Hence, the eigenvalues $\lambda_i$ are non-zero for each $i$.

Let us now consider a function $f$ satisfying $D_{\bar{z}}f = 0$ on the boundary. We may mode expand $D_{\bar{z}}f$ as
\begin{equation}
D_{\bar{z}}f = \sum_{i=0}^\infty c_if^i_{\bar{z}},
\end{equation}
for some complex coefficients $\{c_i\}$. We will now prove the existence of a function $h^i$ on $HS^2$ with $D_{\bar{z}}h^i = f^i_{\bar{z}}$ for each $i$. Firstly, let $X^0$ be a non-vanishing gauge-covariant holomorphic function ($D_{\bar{z}}X^0 = 0$) on $HS^2$ and let us define $\tilde{h}^i = h^i/X^0$ and $\tilde{f}^i_{\bar{z}} = f^i_{\bar{z}}/X^0$. In terms of $\tilde{h}^i$ and $\tilde{f}^i_{\bar{z}}$, the equation $D_{\bar{z}}h^i = f^i_{\bar{z}}$ becomes $\nabla_{\bar{z}}\tilde{h}^i = \tilde{f}^i_{\bar{z}}$ where $\nabla_{\bar{z}}$ is now the ordinary derivative in a local orthonormal frame. Using an isothermal coordinate $w$ on $HS^2$ (see section \ref{sec:hem_geom}), we may further rewrite the equation $\nabla_{\bar{z}}\tilde{h}^i = \tilde{f}^i_{\bar{z}}$ as $\partial_{\bar{w}}\tilde{h}^i = \tilde{f}^i_{\bar{w}}$ where $\tilde{f}^i_{\bar{w}}$ is the coefficient of the $(0,1)$ form $f^i$ in the $w$ coordinate system and $\partial_{\bar{w}}$ is now an ordinary partial derivative with respect to the coordinate $\bar{w}$. This equation may be integrated by
\begin{equation}
\tilde{h}^i(\hat{w}) = \int_\C \frac{\tilde{f}^i_{\bar{w}}(w)}{\pi(\hat{w}-w)}d\bar{w}dw.
\end{equation}
where $\tilde{f}^i_{\bar{w}}$ has been extended smoothly to a compactly supported function on $\mathbb{C}$.
We have thus proved the existence of a function $h^i$ satisfying $D_{\bar{z}}h^i = f^i_{\bar{z}}$. Now, the definition of $f^i_{\bar{z}}$ implies
\begin{equation}
-4D_{\bar{z}}D_zD_{\bar{z}}h^i = \lambda_iD_{\bar{z}}h^i,
\end{equation}
and so $-4D_zD_{\bar{z}} h^i - \lambda_ih^i$ is a gauge-covariant holomorphic function. Since the eigenvalue $\lambda_i$ is non-zero, we may redefine $h^i$ by adding a gauge-covariant holomorphic function such that $h^i$ becomes an eigenfunction of the Laplacian $-4D_zD_{\bar{z}} h^i = \lambda_ih^i$. Finally, we have that $f - \sum_i c_ih^i$ is a gauge-covariant holomorphic function and can therefore be expanded in an orthonormal basis of gauge-covariant holomorphic functions $X^0,X^0 w,X^0 w^2,\ldots$ lying in the kernel of $-4D_zD_{\bar{z}}$. Together with $\{h^i\}$, these functions form an orthonormal basis of eigenfunctions of the Laplacian $-4D_zD_{\bar{z}}$ satisfying the Neumann boundary conditions.

\section{One-loop determinant as an equivariant index}\label{ap:equivariant_index}

In this appendix we explain an alternative way of computing the one-loop determinants by rewriting them as an equivariant index of $Q^2$, where $Q$ is the localising supercharge. The first half of this section is a review of the arguments presented in \cite{fujitsuka2014higgs}.

Let us assume that we have a $Q$-exact localising action $S = QV$ and we seek to compute a one loop determinant as a Gaussian path integral
\begin{equation}
\int D\phi \exp(-\delta^{(2)}S(\phi)) = \int D\phi \exp(-Q\delta^{(2)}V(\phi)),
\end{equation}
where $\delta^{(2)}$ denotes quadratic fluctuations around the BPS locus.

As an illustrative example, for fields $\phi_1,\phi_2,\phi_3$ and localising action $V = \phi_1\phi_2\phi_3$ the quadratic fluctuation is
\begin{equation}
\delta^{(2)}V = \delta\phi_1 \delta \phi_2 \phi_3 + \phi_1\delta\phi_2\delta\phi_3 + \delta\phi_1\phi_2\delta\phi_3.
\end{equation}
Keeping first order terms, the action of $Q$ on the space of fluctuations $\delta \phi_i$ is defined by $Q\delta\phi_i = \delta (Q\phi_i)$. An example relevant to the context of the present work is
\begin{equation}
Q\psi = (\nabla_3 + iA_3 - \sigma)X \quad \text{implies}\quad Q\delta\psi = (\nabla_3 + iA_3 - \sigma)\delta X + (i\delta A_3 - \delta \sigma)X.
\end{equation}
The action of $Q$ on the elementary fields $\phi_i$ is the same as before. For example, on $\delta^{(2)}V$:
\begin{equation}
Q(\delta \phi_1\delta \phi_2 \phi_3) = (Q \delta\phi_1) \delta\phi_2 \phi_3 + \delta \phi_1 (Q\delta \phi_2) \phi_3 + \delta\phi_1\delta\phi_2 (Q\phi_3)
\end{equation}
Note that not all of the terms above contribute to $\delta^{(2)} S$. For example, the term $\delta \phi_1\delta\phi_2 (Q\phi_3)$ does not contribute because $Q\phi_3 = 0$ at the BPS locus by definition. Now, $\delta \phi_1$ and $\delta \phi_2$ must have opposite Grassmann parity because $\phi_3$ needs to be even in order to be nonzero. If there is no $\phi_3$ field then, since $V$ is odd, $\delta \phi_1$ and $\delta \phi_2$ must have opposite Grassmann parity.

By analogy with the above $\delta^{(2)}V$ can be in general written as
\begin{align}
\delta^{(2)}V = 
\begin{pmatrix}
\delta \chi_1& \delta \chi_2 &...
\end{pmatrix}
\begin{pmatrix}
\vdots &\vdots&\vdots\\
\vdots &\vdots&\vdots\\
\vdots &\vdots&\vdots\\
\end{pmatrix}
\begin{pmatrix}
\delta \phi_1\\
\delta \phi_2\\
\vdots
\end{pmatrix}
\end{align}
where $\delta \chi_i$ are the fluctuation of fermionic elementary fields and $\delta \phi_i$ are the fluctuation of bosonic elementary fields. The matrix consists of differential operators depending only on the background field at the BPS locus.
 In some cases (and certainly for all the examples in this paper) we have
 \begin{align}
\delta^{(2)}V = 
\begin{pmatrix}
Q\delta \bar{X}_B&\delta  \bar{X}_F
\end{pmatrix}
\begin{pmatrix}
D_{00} &D_{01}\\
D_{10} &D_{11}\\
\end{pmatrix}
\begin{pmatrix}
\delta X_B\\
Q \delta X_F
\end{pmatrix},
\end{align}
where $X_B$ denotes a collection of bosonic elementary fields and $X_F$ denotes a collection of fermionic elementary fields. Here, $\bar{X}_B, \bar{X}_F$ are complex conjugates, and  $D_{00}, D_{01},D_{10},D_{11}$ are first order differential operators depending only on the background BPS configuration.

We can now compute the quadratic fluctuation $Q\delta^{(2)}V$:
\begin{align}
Q\delta^{(2)}=  \delta^{(2)}V = &
\begin{pmatrix}
\delta \bar{X}_B&Q\delta \bar{X}_F 
\end{pmatrix}
\begin{pmatrix}
Q^2 & 0\\
0 & 1
\end{pmatrix}
\begin{pmatrix}
D_{00} &D_{01}\\
D_{10} &D_{11}\\
\end{pmatrix}
\begin{pmatrix}
\delta X_B\\
Q \delta X_F
\end{pmatrix}\\
&+\begin{pmatrix}
Q\delta \bar{X}_B&\delta \bar{X}_F 
\end{pmatrix}
\begin{pmatrix}
D_{00} &D_{01}\\
D_{10} &D_{11}\\
\end{pmatrix}
\begin{pmatrix}
1 & 0\\
 0 & Q^2
\end{pmatrix}
\begin{pmatrix}
Q\delta X_B\\
 \delta X_F
\end{pmatrix},
\end{align}
where we have used the fact that $Q$ only acts on the fluctuations of the fields but not on the BPS locus. The first line is the bosonic part of the quadratic fluctuation and the second part is the fermionic part of the fluctuation. 

In conclusion the one loop determinant is
\begin{equation}
Z_{1\text{-loop}} = \frac{
\det\left(\begin{pmatrix}
D_{00} &D_{01}\\
D_{10} &D_{11}\\
\end{pmatrix}
\begin{pmatrix}
1 & 0 \\
0 & Q^2
\end{pmatrix}
\right)}{\det\left(
\begin{pmatrix}
Q^2 & 0\\
0 & 1
\end{pmatrix}
\begin{pmatrix}
D_{00} &D_{01}\\
D_{10} &D_{11}\\
\end{pmatrix}\right)}
=\frac{\det\begin{pmatrix}
1 & 0\\
0 & Q^2
\end{pmatrix}}{\det\begin{pmatrix}
Q^2 & 0\\
0 & 1
\end{pmatrix}} = \frac{\det(Q^2)_{\text{Coker} D_{10}}}{\det(Q^2)_{\text{Ker} D_{10}}}.
\end{equation}

In the first equality we have assumed that $\delta X_B$ and $Q\delta X_F$ can be treated as independent variables in the path integral. Let us now return to the context of the main body of this work and compute the one-loop determinant of a Neumann chiral in a monopole background using this formalism. We assume that the real masses are set to zero for simplicity. In this case $V$ is given by
\begin{equation}
(\nabla_3 + iA_3 + iA_3^{(B)} - \sigma)X \bar{\psi} + (\nabla_{\bar{z}} + iA_{\bar{z}})X \bar{\psi}_z + \bar{\psi} \sigma X,
\end{equation}
and we choose $X_B = X, \bar{X}_B = \bar{X},X_F = \psi_{\bar{z}}$ and $\bar{X}_F = \bar{\psi}_z$. The differential operator is given by
\begin{equation}
\begin{pmatrix}
D_{00}&D_{01}\\
D_{10}&D_{11}
\end{pmatrix}
=
\begin{pmatrix}
\nabla_3 + iA_3 + iA_3^{(B)} + \sigma &0\\
(\nabla_z + iA_z)&1
\end{pmatrix}
\end{equation}
where we have assumed that the fluctuation of the gauge multiplet is zero. One can show that $Q^2 = \nabla_3 + iA_3 + iA_3^{(B)} - \sigma$ when acting the on the chiral multiplet fluctuations and so in conclusion the one-loop chiral multiplet determinant is
\begin{equation}
\frac{\det(\nabla_3 + iA_3 +iA_3^{(B)}- \sigma)_{\text{Coker}{\nabla_{\bar{z}} + iA_{\bar{z}}}}}{\det(\nabla_3 + iA_3 + iA_3^{(B)} - \sigma)_{\text{Ker}{\nabla_{\bar{z}} + iA_{\bar{z}}}}}.
\end{equation}
This reproduces the results in equations \eqref{dirichlet_one_loop} and \eqref{neumann_one-loop}.

\bibliographystyle{JHEP}
\bibliography{higgs}

\end{document}